\documentclass[aps,prx,longbibliography,twocolumn,superscriptaddress]{revtex4-2}
\usepackage{setspace}
\usepackage{verbatim}
\usepackage{lmodern}
\usepackage{siunitx}
\usepackage{etoolbox}
\usepackage{booktabs}
\usepackage{tabularx}
\usepackage{makecell}
\usepackage{graphicx}
\usepackage{dcolumn}
\usepackage{bm}
\usepackage{bbm}
\usepackage{bbold}
\usepackage{physics}
\usepackage{enumerate}
\usepackage[dvipsnames]{xcolor}
\usepackage{dsfont}
\usepackage{times}
\definecolor{darkblue}{HTML}{023e8a}
\definecolor{darkred}{HTML}{780000}
\definecolor{darkgreen}{HTML}{005f73}
\usepackage[colorlinks=true,
	    urlcolor=darkred,
	    citecolor=darkblue,
	    linkcolor=darkred
            ]{hyperref}

\begin{document}

\title{Restart uncertainty relation for monitored quantum dynamics}

\author{Ruoyu Yin}
\email{yinruoy@biu.ac.il}
\author{Qingyuan Wang}
\email{qingwqy@gmail.com}
\thanks{equal contribution to this work.}
\affiliation{Department of Physics, 
Institute of Nanotechnology and Advanced Materials, 
Bar-Ilan University, Ramat-Gan 52900, Israel}

\author{Sabine Tornow}
\email{sabine.tornow@unibw.de}
\affiliation{Research Institute CODE, University of the Bundeswehr Munich, 81739 Munich, Germany} 
 
\author{Eli Barkai}
\email{Eli.Barkai@biu.ac.il}
\affiliation{Department of Physics, 
Institute of Nanotechnology and Advanced Materials, 
Bar-Ilan University, Ramat-Gan 52900, Israel}

\begin{abstract}
We introduce a new time-energy uncertainty relation within the context of restarts in monitored quantum dynamics.
%
Previous studies have established that the mean recurrence time, 
which represents the time taken to return to the initial state, 
is quantized as an integer multiple of the sampling time, 
displaying pointwise discontinuous transitions at resonances.
Our findings demonstrate that,
the natural utilization of the restart mechanism in laboratory experiments, 
driven by finite data collection time spans, 
leads to a broadening effect on the transitions of the mean recurrence time.
Our newly proposed uncertainty relation captures the underlying essence of these phenomena,
by connecting the broadening of the mean hitting time near resonances, 
to the intrinsic energies of the quantum system 
and to the fluctuations of recurrence time.
Our uncertainty relation has also been validated 
through remote experiments conducted on an IBM quantum computer.
This work not only contributes to our understanding of fundamental aspects
related to quantum measurements and dynamics, 
but also offers practical insights for the design of efficient quantum algorithms 
with mid-circuit measurements.
\end{abstract}

\maketitle
The concept of restarting a process is a ubiquitous phenomenon across various disciplines \cite{Evans2011,Majumdar2011}.
When faced with a setback in reaching a desired goal, 
the instinct to restart the process often arises, driven by the hope of achieving better success in subsequent attempts. This notion of restarting, or ``resetting'', gives rise to a compelling paradigm in the realm of classical stochastic processes
\cite{Pal2015,Pal2016,Eule2016,Shlomi2017,Belan2018,Evans2018,Maso2019,Pal2019,Evans2020,Tal2020,Redner2020,Gupta2020,Gupta2022,WangWei2022,Bruyne2022,Pal2023}.
Diffusion processes with resets are the best-studied example \cite{Majumdar2011}. 
In this scenario, a particle undergoes random diffusion but, at periodic or random intervals, 
is brought back to its initial position.
Additionally, within this framework, a specific target awaits the particle's arrival, prompting us to inquire about the time it takes for the particle to reach this target for the first time. 
This random time, both with and without the restart mechanism, is commonly known as the ``first passage time'' and has garnered widespread attention \cite{Redner2001}.
In particular, the notion of restarts plays a pivotal role in expediting search processes, 
making these ideas highly relevant and applicable across diverse fields, 
including biology \cite{Hopfield1974}, computer science \cite{Luby1993,Gomes1998}, 
animal foraging \cite{Denis2014,Boyer2018,Pal2020}, 
the study of chemical reactions \cite{Bel2009,Reuveni2014,Blumer2024}, 
and quantum dynamics 
\cite{Majumdar2018,Rose2018,Belan2020,riera2020measurement,Perfetto2021,Perfetto2021a,Turkeshi2021,das2022quantum,Magoni2022,Ruoyu2023,yin2023b,Modak2023nonhermitian,gupta2023tight,Majumdar2023a,Majumdar2023b,chatterjee2023quest,puente2023,Liu2023Semi}, 
among others.

The concept of restarting processes is of particular importance 
in the context of repeated mid-circuit measurements performed on quantum computers 
and more generally in the context of monitored quantum walks \cite{Gruenbaum2013}.
In quantum dynamics, the notion of ``first hitting time'' without restart 
reveals intriguing and novel features,
often intimately connected with topological considerations, resonances, 
and the concept of dark states
\cite{Krovi2006,Gruenbaum2013,Dhar2015a,Dhar2015,Friedman2017a,Felix2018,Silberhorn2018,Lahiri2019,yin2019,Thiel2020D,Thiel2020,Parker2020,Thiel2021entropy,David2021,Ziegler2021,dubey2021quantum,Didi2022,Liu2022a,Das2022,Yajing2023,Zhenbo2023,walter2023thermodynamic,Sabine2022,Laneve2023,Wang2024Entropy}. 
Typically, these processes are represented using graphs, 
which can describe the states of various quantum systems, 
such as single particles or qubit systems.
Within this graph, a crucial element is the presence of a target state,
often symbolizing the measurement device.

To detect the system at the target state, 
it might be tempting to perform measurements at infinitesimally short intervals.
However, this approach encounters the Zeno effect \cite{Misra1977}, 
where frequent strong measurements effectively freeze the system's dynamics, 
rendering it undetectable. 
As a solution, a sequence of measurements is performed 
at regular intervals of $\tau$ units of time,
allowing the system to evolve unitarily between measurements 
\cite{Krovi2006,Gruenbaum2013,Dhar2015a,Dhar2015,Friedman2017a,Felix2018,Silberhorn2018}.
Yet, when implementing this fundamental search process on a quantum computer 
or any practical device, practical challenges emerge. 
Over time, due to measurement imperfections or interactions with the environment, 
quantum effects tend to diminish due to noise and decoherence. 
In such cases, a common strategy is to restart the process. 
This issue of finite-time resolution is not exclusive to the quantum realm 
and is encountered in classical systems as well.
What distinguishes the quantum realm is the potential for sharp and discontinuous resonances 
in mean hitting times, related to quantum revivals \cite{jayakody2022} 
and topological effects (see below).
Remarkably, as shown below, even when the restart time is significantly longer than 
the mean first hitting time, the act of restarting can have a profound impact.
Our objective is to investigate these phenomena by leveraging a new uncertainty relation,
which is vastly different from previous ones 
\cite{Gingrich2017,Garrahan2017,Falasco2020,Pal2021prr,Hasegawa2022,bebon2023}.

\begin{figure*}
\centering
\includegraphics[width=11.4cm]{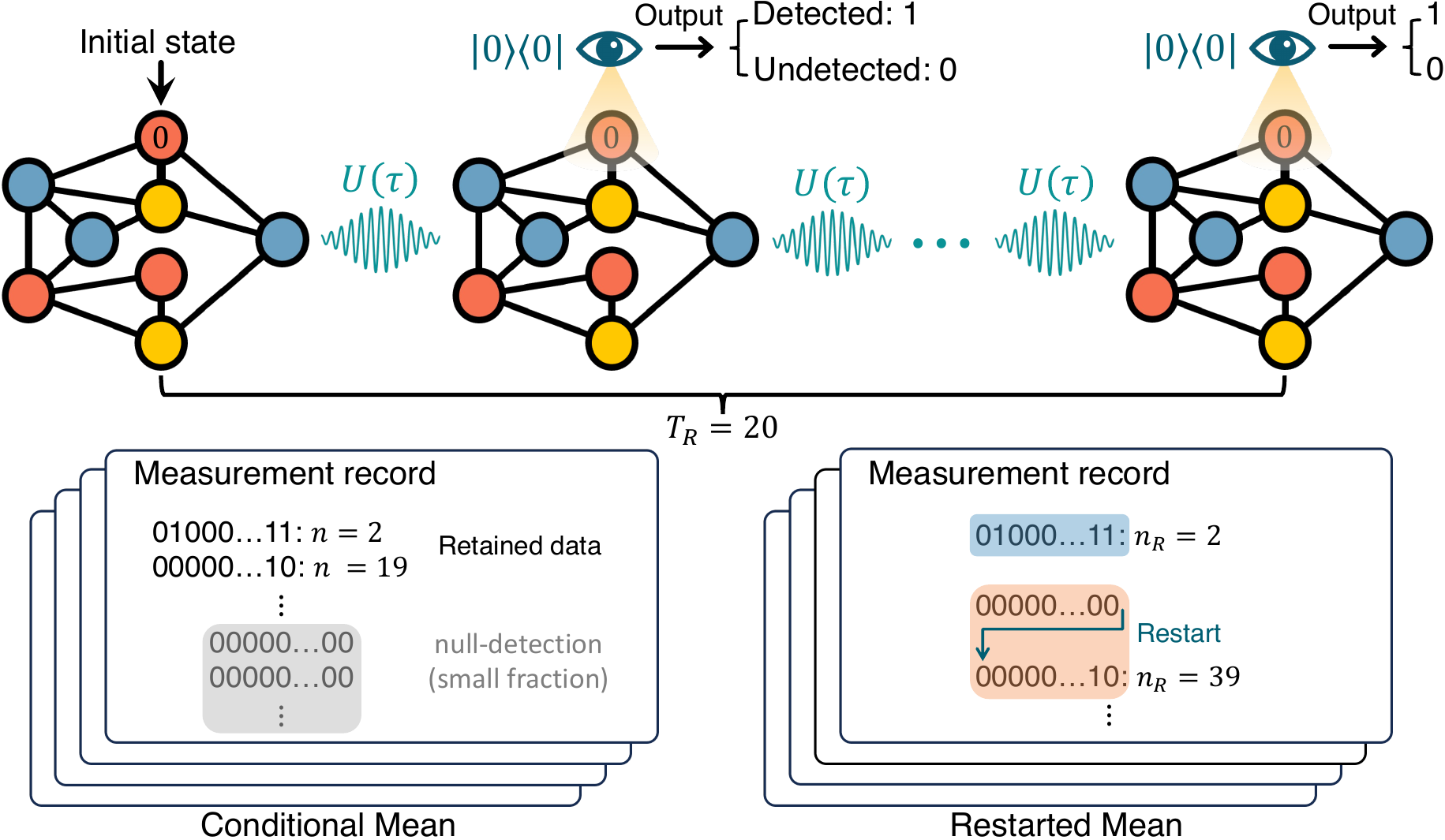}
\caption{
{The measurement protocol for monitored quantum walks and its output.}
A quantum walker on a graph is initialized at the spatial state $\ket{0}$ (marked ``0'').
A projective measurement at the initial state, schematically presented by the eye symbol, 
is performed following the unitary evolution of time $\tau$.
The output of the measurement is either ``yes'' ($1$) or ``no'' ($0$),
rendering the wavefunction of the quantum walker either localized at $\ket{0}$
or its amplitude erased at the state $\ket{0}$.
We continue the free evolution immediately after the measurement for another duration $\tau$,
and then measure again, resulting in another binary outcome: $0$ or $1$.
Using an IBM quantum computer, the process of interrupting evolution 
by stroboscopic measurements, {for a tight-binding three-site ring}, 
was implemented for $20$ steps, as a single realization, 
thus leading to an output string or measurement record of $20$ bits. 
Our goal is to find the number of steps when the first $1$ (``yes'') emerges, 
which is the quantum first hitting time in units of $\tau$.
Repeating a large number of realizations gives the statistics of hitting times. 
Two common statistical measures of estimating the mean hitting time are used.
In the first we disregard the (rare) sequences with all 0 measurements, and this yields the mean conditioned on detection.
In the second, called restarted hitting time, we continue until the first detection, as illustrated in the figure, leading to the sampling of the mean restarted hitting time.
In this example the restart time is $T_R =20$ in units of $\tau$. 
}
\label{fig:protocol}
\end{figure*}

To illustrate the key aspects of our study, 
we commence with an experimental demonstration conducted on an IBM quantum computer. 
In this experiment, we consider a straightforward three-site ring graph 
with quantum states represented as $\ket{0}$, $\ket{1}$, and $\ket{2}$. 
The system is described by a tight-binding Hamiltonian 
that accounts for hopping between these states.
Our starting point is state $\ket{0}$, 
which also serves as the target state for this investigation. 
We aim to observe the recurrence of the system to its initial state 
through periodic measurements 
conducted every $\tau$ unit of time. 
The measurement outcomes yield a sequence of ``no'' responses (indicating null detection) 
followed by a ``yes'' response when the target state is eventually detected. 
The first occurrence of ``yes'' in this sequence defines the first hitting time 
\cite{Krovi2006,Gruenbaum2013,Dhar2015a,Dhar2015,Friedman2017a,Felix2018,Silberhorn2018}, 
as demonstrated in Figure \ref{fig:protocol}. 
For instance, an experimental outcome might yield the sequence 
$\{no, no, yes\}$, which corresponds to a first detection time of $3 \tau$.
Through repeated experiments conducted on the quantum computer, 
we determine the mean number of measurements required for detection, 
denoted as $\langle n \rangle$. 
This quantity, extracted from the quantum computer, 
provides us with valuable insight into the average time it takes to detect the target state.

Theoretical investigations, spanning a wide range of graph types, 
have extensively explored the aforementioned problem
\cite{Krovi2006,Gruenbaum2013,Dhar2015a,Dhar2015,Friedman2017a,Felix2018,Silberhorn2018,Lahiri2019,yin2019,Thiel2020D,Thiel2020,Thiel2021entropy,David2021,Ziegler2021,dubey2021quantum,Yajing2023,Zhenbo2023,walter2023thermodynamic,Parker2020}. 
We first present the basic theory ignoring restart, 
showing that such a theory does not align with the experiments.  
Notably, Grünbaum and colleagues \cite{Gruenbaum2013} made a remarkable discovery: 
The theoretical mean recurrence time exhibits quantization. 
In practical terms,
this implies that the value of $\langle n \rangle$ 
is constrained to integer values. 
Mathematically, this integer is encapsulated by a winding number $w$
associated with a generating function and hence
the phenomenon is topological.  
The integer is defined and denoted as 
\begin{equation}
\langle n \rangle = \sum_{n=1} ^\infty n F_n =  w.
\label{eq01}
\end{equation}
Here $F_n$ is the probability of first detection in the $n$-th measurement,
which is normalized, i.e. $\sum_{n=1}^\infty F_n = 1$.
It is obtained using the unitary $U(\tau)=\exp(- i H \tau)$ 
($\hbar$ is set as $1$, and $H$ is the Hamiltonian)
describing the evolution between measurements 
and the projection $\ket{0}\bra{0}$ describing the measurements using collapse theory, so all along this work $\ket{0}$ is the target state.
Specifically \cite{Gruenbaum2013,Dhar2015a,Friedman2017a},
\begin{equation}
F_n = \left| \bra{0} U(\tau){\cal S}^{n-1} \ket{0} \right|^2,
\label{eq01a}
\end{equation}
where the survival operator ${\cal S} = \left( \mathbbm{1} -\ket{0}\bra{0} \right) U(\tau)$
($\mathbbm{1}$ is the identity matrix),
demonstrating the unitary evolution in the time interval $\tau$ 
followed by the complementary projection described by $\mathbbm{1}-\ket{0}\bra{0}$ 
(indicating null detection). 
In general, the winding number $w$ is computed as follows \cite{Gruenbaum2013,yin2019}:
Given the time-independent Hamiltonian and assuming a finite graph, 
we search for the energy levels and corresponding states of the system,
denoted as $H \ket{E_k} = E_k \ket{E_k}$. 
The value of $\langle n \rangle = w$ 
represents the count of distinct phase factors, such as $e^{-i E_k \tau}$,
associated with stationary states that exhibit nonzero overlap with the target state.
See details including the proof for equation (\ref{eq01}) in the Supplementary Information (SI)
\begin{figure}[htbp]
\centering
\includegraphics[width=0.95\linewidth]{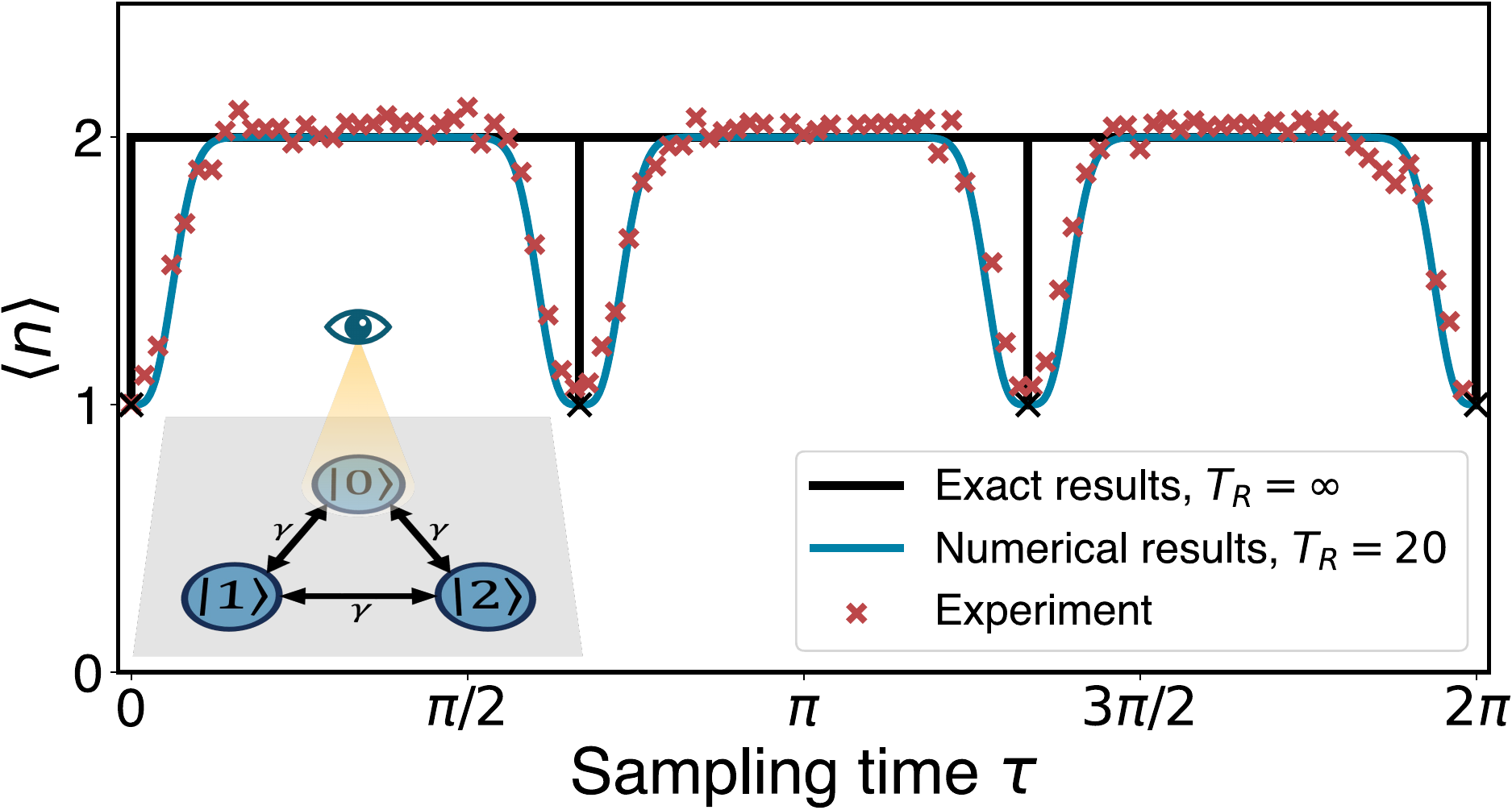}
\caption{{Mean hitting time for the three-site ring model.}
The numerically/experimentally obtained mean quantum first return time of the three-site ring model.
The exact results for $T_R = \infty$ (black line), 
as stated under equation (\ref{eq01}), 
present discontinuous jumps or dips of $\expval{n}=w$,
from $w=2$ to $w=1$, at $\tau=2\pi k /3$ ($k=0,1,2,\dots$). 
In the experimental data (red crosses, $T_R = 20$), these transitions are widened.
The numerical results for $T_R = 20$ (blue line) 
perfectly match the experimental results.
In the paper, we address the broadening effect 
showing how it is related to an uncertainty relation.
Inset is the scheme of the tight-binding model for a ring with three sites,
and $\gamma$ (set as $1$) denotes the strength of the hopping matrix element,  see equation (\ref{eq1}).
We measure periodically the target state $\ket{0}$ (indicated with an eye).
See details of the IBM remote experiments in Materials and Methods, and SI. 
}
\label{fig:Theomean}
\end{figure}

In our experimental example on the three-site ring 
(see Materials and Methods: Model), we encounter energy level degeneracy,
resulting in $\langle n \rangle = 2$ for nearly any choice of $\tau$.
However, a pivotal observation emerges when the phase factors merge,
causing $\langle n \rangle$ to become equal to $1$. 
The merging of phase factors occurs for specific values of $\tau$ 
which are straightforward to identify. 
Consequently, the relationship between $\langle n \rangle$ and $\tau$ 
is predominantly characterized by the value $2$,
except for isolated pointwise discontinuities, where it abruptly becomes $1$.
These peculiar values of $\tau$ correspond to instances of wave packet revivals,
wherein certain times lead to the complete revival of the wave packet to its initial state.
During such moments, the first measurement invariably yields a ``yes'' outcome.
What makes this phenomenon particularly extraordinary
is the discontinuous nature of $\langle n \rangle$ 
and its intriguing insensitivity to values of $\tau$ beyond the revival times themselves.

The theoretical findings described above are valid in principle 
for infinitely long time measurements, 
and they have been graphically represented in Figure~\ref{fig:Theomean}, 
alongside the corresponding experimental results from 
an IBM Eagle processor (IBM Sherbrooke).
Notably, the delta-like narrow transitions predicted by the theory 
are observed to exhibit widening in the real-world experimental data. 
Nonetheless, a clear alignment between theory and experiment persists, 
except in the immediate vicinity of these transitions.
Importantly, the above-mentioned resonances and broadening effect
is a generic phenomenon of first hitting time statistics, 
and is not limited to the example under study.

The inception of this research stemmed from the natural inquiry: 
Is this widening phenomenon a generic occurrence?
Is it primarily attributed to inherent noise inherent to the system,
such as imperfect timing in measurements or the unitary itself 
or is it potentially linked to the fundamental principles of quantum measurement theory?
Specifically, can the basic postulates of quantum measurement theory provide 
a quantitative description of these transitions? 
When we refer to a ``transition'', or a ``topological transition'' or ``resonance'', 
we mean the shift of  $\langle n \rangle = w$ 
(as illustrated by $w=2$ in Figure~\ref{fig:Theomean}) 
to $\langle n \rangle = w-1$ and back, 
as we systematically vary the parameter $\tau$.
In this context, $\tau$ serves as our control parameter,
although it is worth noting that other parameters of the system Hamiltonian 
could be employed for a similar investigation.
We claim below that the widening effects seen in Figure~\ref{fig:Theomean}, 
are generic and are due to the restart paradigm. 
Secondly,
we find that the widening effects are determined by the fluctuations in the system,
or to put it differently, 
the width of the transition teaches us about the fluctuations of the hitting time.
Further, these uncertainties in hitting times are shown 
to be related to the energies of the system, 
thus extending the time-energy uncertainty relation to a case 
where the time is actually fluctuating.
\begin{figure*}[htbp]
\begin{center}
\includegraphics[width=\linewidth]{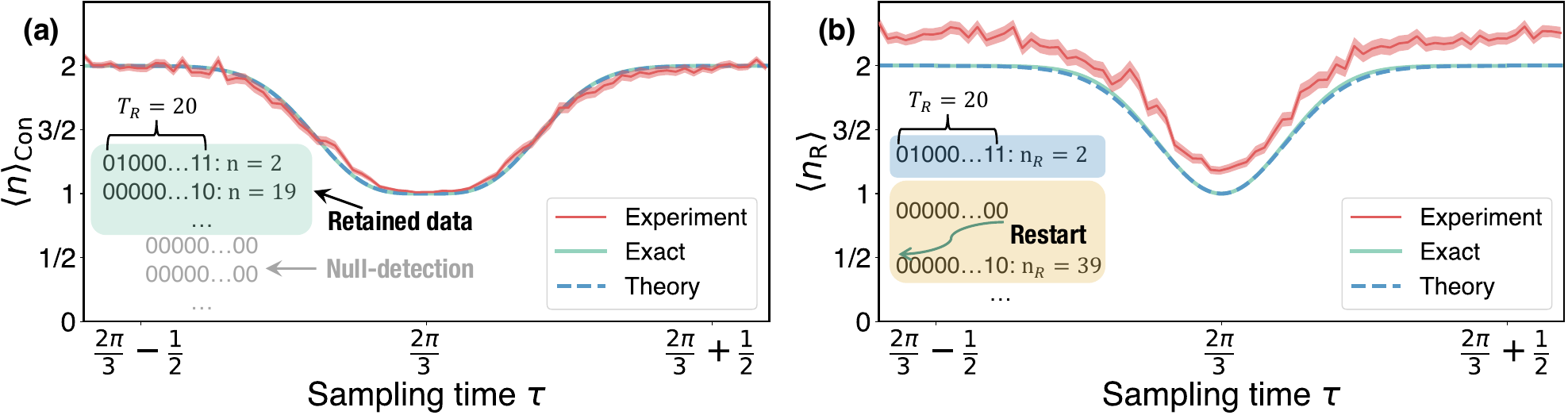}
\end{center}
\caption{{Impact of restart on recurrence time transitions.} 
(a) The transition from $\expval{n}_{\text{Con}} = 2$ to $\expval{n}_{\text{Con}} = 1$ 
and back is widened due to restarts. 
In particular, here we restart after $T_R=20$ measurements, as highlighted in the insets.
We compare the exact results (green solid line) found using equations~(\ref{eq01a},\ref{eq02})
with the theory (blue dashed line) obtained using equation (\ref{eq05}) 
and IBM quantum computer experiments (red line).
The results clearly demonstrate that basic postulates of measurement theory 
and the uncertainty relation using the variance of the hitting time 
perfectly align with the experiment. 
In turn, noise and imperfect measurements are not factors in the observed behaviour. 
(b) The mean hitting time under restart, $\expval{n_R}$, as a function of $\tau$.
We compare the exact results (green solid line), 
the theory (blue dashed line, computed with equation (\ref{eq06})) 
and experiment results on the IBM quantum computer (red line) for $T_R = 20$.
We observe the vertical shift between the experimental and exact results,
which is due to noise in quantum computers,
and more specifically, due to a small $1$\% shift in the detection probability
which is discussed in the text.
The model here is a tight-binding three-site ring, the same as in Figure \ref{fig:Theomean}.
In both figures, the exact results are obtained using equation (\ref{eq01a}), 
from which we find $F_n$, and then using equation (\ref{eq02}) for (a) 
or equation (\ref{eq03}) for (b).
The shaded red region represents the confidence interval $99.7\%$,
signifying an interval spanning three standard deviations above and below
the mean in a standard normal distribution.
}
\label{fig:Tridip}
\end{figure*}

Using mid-circuit measurements, 
the experimental output typically commences with a sequence of null measurements,
characterized by the string $\{no, no, \dots\}$. 
It is important to note that this string is always finite, 
and its length is denoted as $T_R$ (with the subscript ``R'' signifying ``restart''). 
In some instances, we encounter a ``yes'' in the sequence,
signifying the successful detection of interest, 
and thus providing the random hitting time. 
However, there are cases where we find a sequence 
composed entirely of ``no's'', 
implying that no detection has occurred until the time $ T_R \tau $, 
see Figure~\ref{fig:protocol} with $T_R = 20$. 
To analyze the statistical features of the experiments, 
we use basics of restart theory.
When we average the results, we focus on two essential statistical measures.
The first is the mean, conditioned on detection within the first $T_R$ attempts, 
denoted as $\langle n \rangle_{{\rm Con}}$, is given by:
%
\begin{equation}
\langle n \rangle_{{\rm Con}} 
= \frac{\sum_{n=1} ^{T_R} n F_n}{P_{\rm det}},
\label{eq02}
\end{equation}
%
where $P_{\rm det} := \sum_{n=1} ^{T_R} F_n$ 
is defined as the detection probability within time $T_R$.
In the estimation of this mean, we exclude all sequences that contain $T_R$ null measurements.
The second statistical measure is the restarted mean, 
which counts all sequences, including those without any ``yes'',
denoted as $\langle n_R \rangle$.
Namely, $n_R$ gives the total number of attempts until the first ``yes'',
regardless of how many restarts have happened. See the schematics in Figure \ref{fig:protocol}.
Its mean is quantified as \cite{Pal2021,Ruoyu2023}:
\begin{equation}
\begin{split}
\langle n_R \rangle 
=\langle n \rangle_{{\rm Con}} +T_R { {1-P_{\rm det}} \over {P_{\rm det}}}.
\end{split}
\label{eq03}
\end{equation}
The first term on the right-hand side corresponds to paths 
where detection occurred within $T_R$ attempts, 
while the second term encompasses paths where detection happened after $T_R$ attempts. 
Therefore, the mean restart time, $\langle n_R \rangle \tau$, 
provides an estimate of the average time until the first detection, 
considering an ensemble that does not exclude any specific path.
In theory, as $T_R$ tends toward infinity, 
we obtain the idealized limit as expressed in equation (\ref{eq01}) 
from equations (\ref{eq02}) and (\ref{eq03}),
though precisely in the vicinity of resonances, 
this limit must be considered with care.  

We introduce the variance of detection times, measured in units of $\tau$, as:
\begin{equation}
 {\large \sigma}_{n} ^2= \langle n^2 \rangle - \langle n\rangle^2 
 = \sum_{n=1} ^\infty n^2 F_n - w^2.
\label{eq04}
\end{equation}
This variance, denoted as $\sigma_{n} ^{2}$, 
quantifies the uncertainty associated with the first hitting time.
Importantly, this uncertainty tends to be substantial 
in the proximity of the topological transition under investigation, 
and notably, these fluctuations become more pronounced as we approach the transition
\cite{yin2019}.
Our main results are relationships between this uncertainty 
and the restarted process using the following expressions:
\begin{equation}
\langle n \rangle_{{\rm Con}} 
= w - \left( {2T_R \over \sigma_{n} ^2 } + 1 \right) 
\exp \left( -{2T_R\over \sigma_{n} ^2} \right),
\label{eq05}
\end{equation}
\begin{equation}
\langle n_R \rangle = w - \exp\left(- {2 T_R \over \sigma_{n} ^2} \right).
\label{eq06}
\end{equation}
These equations hold in the limit of large $T_R$ 
and large $\sigma_{n}^2$ while keeping the ratio $T_R/\sigma_{n} ^2$ constant.
These relationships are general in nature,
describing transitions from $w$ to $w-1$, 
a phenomenon found in a broad class of Hamiltonians when a pair of phase factors merge.
When $T_R/\sigma_{n} ^2 \to \infty$, signifying a state far from resonance, 
we observe that $\langle n \rangle_{{\rm Con}} = \langle n_R \rangle = w$.
Conversely, when $T_R/\sigma_{n} ^2 \to 0$, indicating resonance, 
we find that $\langle n \rangle_{{\rm Con}} = \langle n_R \rangle = w-1$.
Thus, equations (\ref{eq05},\ref{eq06}) describe the broadening of the transitions
that diminishes as we increase the resetting time. 
These findings are significant as many aspects of the process, such as the complete spectrum of $\cal S$ or $U$, are unimportant and do not impact the overall outcome.
We will soon show that this is related to a new type of time-energy relation.


\section*{Experimental Validation}
In the analysis of the experimental data depicted in Figure~\ref{fig:Tridip}(a), 
we relied on the use of the conditional mean, as described earlier.
Additionally, we provided a theoretical representation based on equation (\ref{eq05}), 
which exhibits a remarkable alignment with the experimental 
results without requiring any fitting procedures. 
This indicates that the uncertainty relation, 
solely based on measurement postulates and not noise in the IBM quantum computer, 
is responsible for the broadening.
For these experiments, we set $T_R=20$.
%
Interestingly, in Figure~\ref{fig:Tridip}(b), 
for the restarted mean, we also observe an alignment of the theory with experiment,
though now we see a small constant shift between predictions and the data.
We now explain this effect.

Consider $\tau$ in Figure~\ref{fig:Tridip}(b) far from resonance,
for instance, at $\tau = 2 \pi / 3$,
the theoretical detection probability within time $T_R = 20$, 
$P_{\rm det}=\sum_{n=1}^{T_R} F_n$, 
is approximately equal to $1$. 
However, in our experimental observations, 
we find that $P_{{\rm det}}$ is approximately $0.99$, 
indicating a small but notable deviation between theory and experiment.
This slight deviation has a noticeable impact on the 
expected value of $ n_R $.
Recall that $(1 - P_{{\rm det}})T_R/P_{{\rm det}}$, 
i.e. the second term in equation (\ref{eq03}), is approximately $0$,
since $P_{{\rm det}} \simeq 1$. 
However, when we use the experimental values just mentioned,
we find that for $T_R = 20$, $T_R(1-P_{\text{det}})/P_{\text{det}} = 0.2$.  
Remarkably, this observed value corresponds exactly to 
the shift we observe in $\langle n_R \rangle$, as presented in Figure~\ref{fig:Tridip}(b) 
(please refer to the Supplementary Note 1 in SI for an in-depth discussion on this issue).
We conclude that the small shift is consistent with very small errors 
in the estimation of the detection probability $P_\text{det}$.

%

This situation highlights a crucial point: 
when $T_R$ is large, 
even small errors on the order of $1$\% 
can result in a visible shift in the experimental outcome, 
$\langle n_R \rangle$, and this shift grows linearly with $T_R$.
A similar effect is not found for the conditional mean.
As mentioned, the latter neglects experimental realizations with no detection at all.
The conditional mean consistently falls below the restarted mean, 
a trend particularly noteworthy in search contexts, 
where the primary objective is to expedite the process.
Hence one should wonder which measure holds greater merit.
We believe that both are valuable statistical measures, 
and there is no point in highlighting one over the other.
We will later address the noise issue in our experiment, 
now we return to the theoretical analysis of the uncertainty relation.
\begin{figure*}[htbp]
\begin{center}
\includegraphics[width=\linewidth]{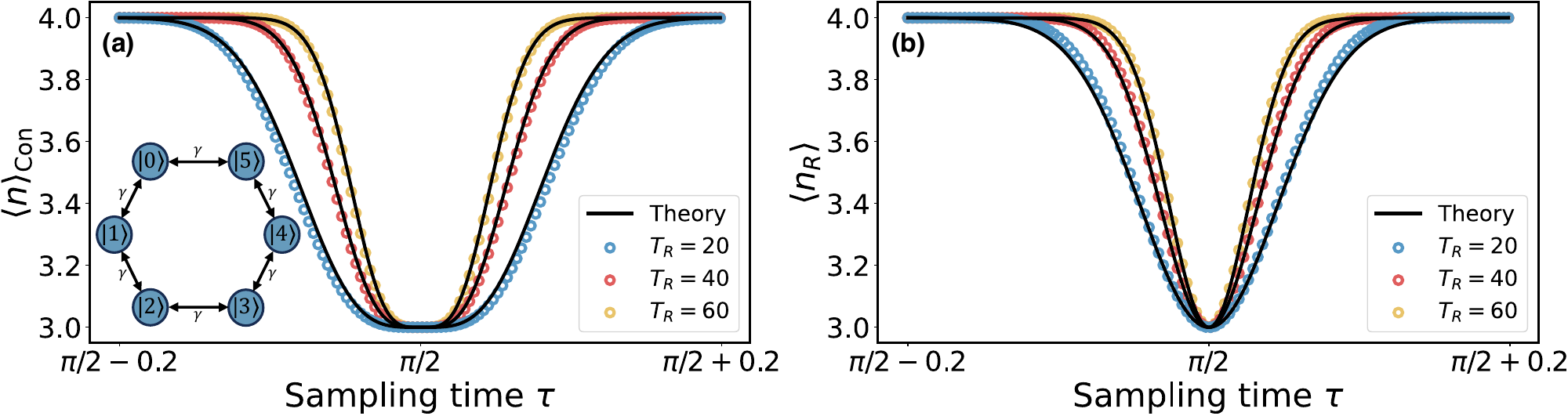}
\end{center}
\caption{{The broadening of the recurrence time transitions in the benzene-type ring model.} 
(a) The conditional mean $\expval{n}_\text{Con}$ and (c) the restart mean $\expval{n_R}$
as a function of $\tau$. 
The model here is the benzene-type ring (equation (\ref{eq1}) with $L=6$ and $\gamma = 1$),
and we work in the vicinity of its critical sampling time $\tau = \pi/2$,
with the transition $\langle n \rangle =4$ to $\langle n \rangle =3$.
The black lines represent the theory from equations~(\ref{eq05}) and (\ref{eq06}).
The dots represent the numerical exact results obtained using equation (\ref{eq01a}).
In the figures, from the bottom to the top line, 
the restart time $T_R$ is $20$, $40$, and $60$, respectively.
Clearly, the transition is narrowed when $T_R$ grows. 
Inset is the scheme of the benzene-type ring model, and the target state is $\ket{0}$.
}
\label{fig:sixcomb}
\end{figure*}
%

\section*{Uncertainty and Energy}
Given that the merging energy phase factors, denoted $\exp(- i E_+ \tau) $ and $ \exp(-i E_- \tau)$, 
are responsible for the resonances observed, 
we aim to establish a connection between the restarted and conditional means 
and the underlying energies within the system.
To accomplish this, 
we provide a sketch of the proof of the main results and extend them.
In the limit of a large number of attempts (denoted as $n$), 
the probability of detection in the $n$-th attempt exhibits exponential decay, 
as expressed by:
\begin{equation}
F_n \sim a(\zeta_\text{max}) \left| \zeta_\text{max} \right|^{2 n}.
\label{eq07}
\end{equation}
$|\zeta_\text{max}|$ is the largest eigenvalue of the survival operator ${\cal S}$ satisfying $|\zeta_\text{max}|<1$. 
$a(\zeta_\text{max})$ is a coefficient independent of $n$ (which will soon be discussed).
A critical aspect to consider is that when we precisely tune $\tau$ to the resonance, 
$|\zeta_\text{max}|\to 1$ (see below for graphic explanation) \cite{Gruenbaum2013,yin2019,Thiel2020D,Liu2022a}.
As we soon explain at resonance $\lim_{|\zeta_\text{max}| \to 1} a(\zeta_\text{max})=0$. 
This occurrence effectively reduces the dimension of the Hilbert space, 
and this reduction can be demonstrated as the reason for the transition 
from $w$ to $w-1$ \cite{Gruenbaum2013}, 
which, in turn, translates to the resonance observed in the hitting time.
To gain insight, let us consider a scenario in which two phase factors have exactly merged, specifically when $\exp(-i E_- \tau) = \exp(-i E_+ \tau)$ for some pair of energy levels. 
In this case, the following state is called dark \cite{Thiel2020D,Liu2022a}:
\begin{equation}
\ket{D} = N\left[ \bra{0} E_+\rangle \ket{ E_-} - \bra{0} E_- \rangle\ket{E_+} \right].
\label{eqdark}
\end{equation}
Here $N$ is for normalization, and ${\cal S} \ket{D} = e^{-i E_+ \tau} \ket{D}$, 
indicating that the eigenvalue of ${\cal S}$ resides on the unit circle. 
Since this state is orthogonal to the target state $\ket{0}$
and also an eigenstate of the unitary,
if we initially populate this state, it is never detected, so it is a dark state.
Hence in our problem, when we adjust the parameter $\tau$, 
which is the focus of our resonance and broadening study, 
we find that it is intricately linked to the creation of a dark state 
within the Hilbert space.
Further, when the parameters are set close to resonance, $|\zeta_\text{max}|$ is close to unity, 
indicating a very slow relaxation of $F_n$, 
which in turn is responsible for the novel effects of the restarted process.

To continue consider the sum in the numerator of equation (\ref{eq02})
using equations (\ref{eq01},\ref{eq07}) 
\begin{equation}
\begin{aligned}
    &\sum_{n=1} ^{T_R}  n F_n  = w - \sum_{T_R} ^\infty n F_n \\
    \sim& 
    \,w - a(\zeta_\text{max})  
    { 
    T_R \left( 1 - |\zeta_\text{max}|^2 \right) + 1
    \over 
    \left( 1 - |\zeta_\text{max}|^2 \right)^2 
    }
    \left| \zeta_\text{max} \right|^{2 (1 + T_R)}
    ,
\end{aligned}
\label{eq08}
\end{equation}
where we summed an infinite series. 
As mentioned when phase factors match, 
the right-hand side of equation (\ref{eq08}), 
based on the theorem in Ref. \cite{Gruenbaum2013}, 
must be $w-1$, when $T_R$ is large. 
It then follows that, 
taking the limit $|\zeta_\text{max}|\to 1$ before $T_R \to \infty$ in equation (\ref{eq08}), 
we find $a(\zeta_\text{max})\sim (1 - |\zeta_\text{max}|^2)^2$, 
a result that can be reached with rigorous arguments.
Applying a similar procedure to the denominator of equation (\ref{eq02}) 
and to equation (\ref{eq03}) leads to the following main result:
let $\rho= T_R (1- |\zeta_\text{max}|^2 )$,
when $|\zeta_\text{max}| \to 1$ and $T_R \to \infty$, 
we find
\begin{equation}
    \langle n \rangle_{{\rm Con}} = w - (\rho+1) e^{-\rho} 
    \ \mbox{and} \ 
    \langle n_R \rangle = \langle n \rangle_{{\rm Con}}  + \rho e^{-\rho}. 
\label{eq09}
\end{equation} 
These formulas relate the resonances and the broadening 
to both the slowest decaying channel in the problem, i.e. to the eigenvalue $\zeta_\text{max}$,
and the restart time $T_R$.
They show how an analysis of the spectrum of the survival operator, 
in particular, the finding of its largest eigenvalue $|\zeta_\text{max}|<1$, is crucial for the problem.

We now consider the fluctuations of the hitting time.
Splitting the sum equation (\ref{eq04}) into two, we have
\begin{equation}
\sigma_{n} ^2 = \sum_{n=1} ^{k_c}  (n - w)^2 F_n + \sum_{k_c+1} ^\infty (n-w)^2 F_n.
\label{eq10}
\end{equation} 
Choosing a large value of $k_c$ such that we can use equation (\ref{eq07}), 
summing an infinite series we find \cite{yin2019} 
$\sigma_{n} ^2 \sim 2/(1 - |\zeta_\text{max}|^2)$. 
This quantifies the statement made before: 
the fluctuations are large close to the transition since $|\zeta_\text{max}|\simeq 1$. 
Using this relation between the uncertainty $\sigma_n$ 
and the eigenvalue $\zeta_\text{max}$ we obtain equations (\ref{eq05},\ref{eq06}). 
A rigorous proof, including the validity of equation (\ref{eq07}),  
is provided in the Supplementary Note 2 in SI.

To complete the physical picture, 
namely, connect the resonance width with the energies of the system,
we use the results in \cite{yin2019}.
A perturbation theory, where the small parameter is the small arc on the unit disk, 
connecting the two nearly merging phases $\exp(- i E_- \tau)$ and $\exp(- i E_{+} \tau)$, 
was used to find $\zeta_\text{max}$.  
The results in Ref. \cite{yin2019} gives 
$\left| \zeta_\text{max} \right|^2 \sim 1-\lambda (\widetilde{\Delta E \tau})^2$ 
(parameters soon to be defined).
Then with equation (\ref{eq09}) we find
\begin{equation}\label{eq11}
    \expval{n}_\text{Con} 
    = w - 
    \left[ 1+ \lambda T_R (\widetilde{\Delta E \tau})^2 \right]
    \exp 
    \left[ -\lambda T_R (\widetilde{\Delta E \tau })^2 \right],
\end{equation}
\begin{equation}\label{eq12}
    \expval{n_R} 
    = w - 
    \exp 
    \left[ -\lambda T_R (\widetilde{\Delta E \tau })^2 \right],
\end{equation}
where $\lambda =  {p_+p_- / (p_+ + p_-)^3}$ 
with the overlaps $p_\pm = \sum_l^{g_\pm} \abs{\braket{0}{E_{\pm,l}}}^2$ 
($g_\pm$ is the degeneracy of the energy level $E_\pm$), 
and 
\begin{equation}
\widetilde{\Delta E \tau } := \tau | E_+ - E_- | \mod 2 \pi. 
\end{equation}
%
Equations (\ref{eq11},\ref{eq12}) clearly show 
the dependence of the mean hitting time on the system energies, 
and also practically, are employed to obtain the theoretical results in Figure \ref{fig:Tridip}. 
At resonances, when $\widetilde{\Delta E \tau} =0$, both  $\langle n \rangle_{\text{Con}}$ and $\langle n_R\rangle$ are equal to $w-1$. Additionally,  the resonance width decreases when we increase the restart time, assuming all other parameters remain constant.

\begin{figure}[t]
\begin{center}
\includegraphics[width=0.95\linewidth]{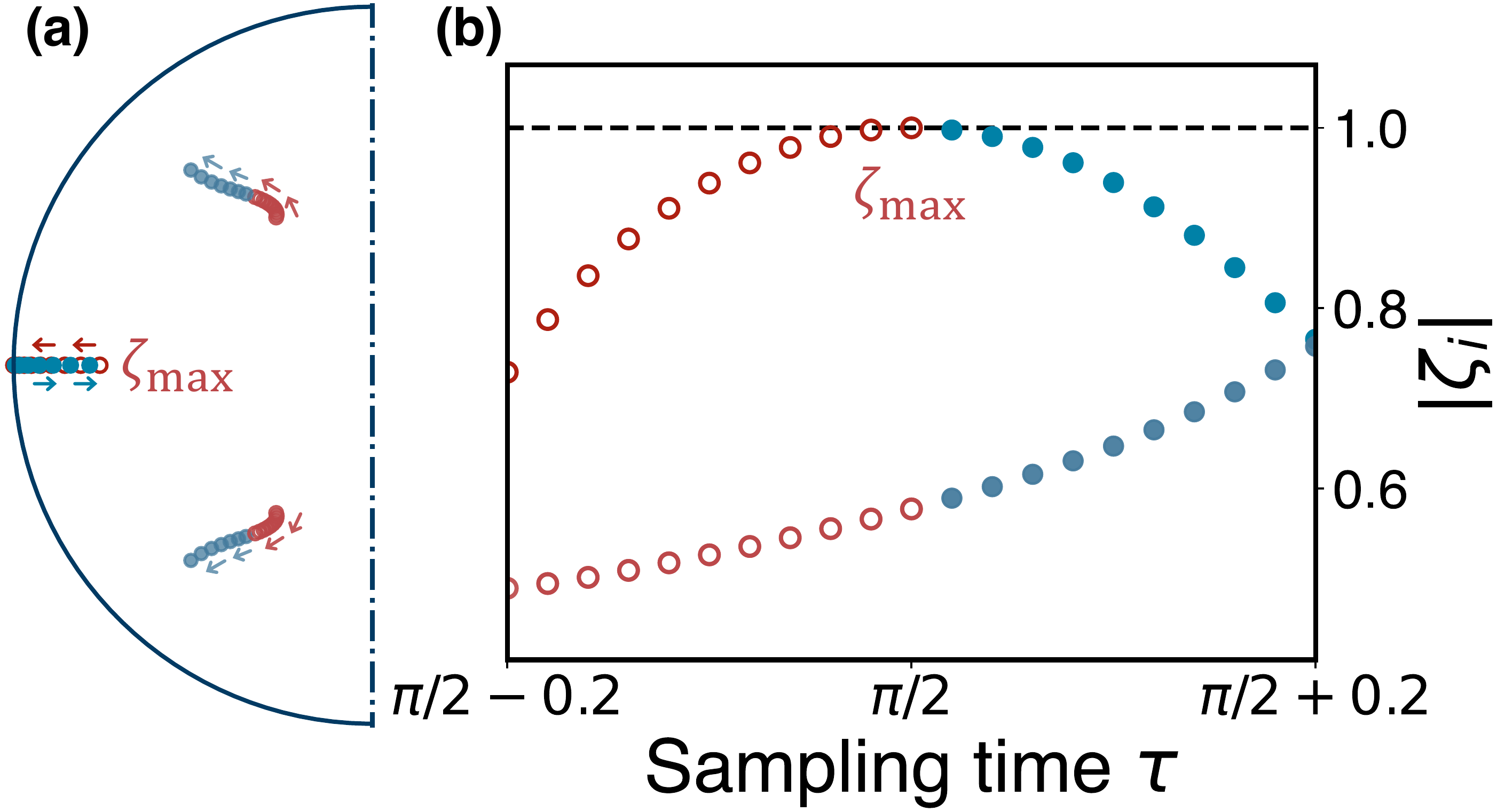}
\end{center}
\caption{{Eigenvalue analysis in the benzene-type ring model.} 
The eigenvalues $\{\zeta_i \}$ of the survival operator ${\cal S}$ for the six-site ring model, 
with the sampling time $\tau$ varied in the same range as in Figure \ref{fig:sixcomb}.
Recall that $\left| \zeta_i \right|$ in general are less or equal unity.
In (a) we present the eigenvalues as the sampling time $\tau$ is varied, 
and the semicircle is of radius $1$.
In (b) we plot the absolute values of $\{\zeta_i \}$.
Due to the degeneracies of ${\cal S}$,
we have three eigenvalues.
As shown in (a), two eigenvalues (conjugate to each other) 
are far away from the unit circle and hence become irrelevant.
One eigenvalue approaches the unit circle, 
and is solely responsible for the hitting time statistics and the uncertainty relation.
We use arrows to illustrate entering or exiting the resonance at $\tau=\pi/2$.
The red open circles present the eigenvalues when entering the resonance,
and blue closed circles are used for the ones when exiting the resonance.
The corresponding behaviors of the distance of the eigenvalues $\{\zeta_i \}$ to the origin 
are demonstrated in (b),
where the two irrelevant eigenvalues share one set of data presented by the lower circles.
Clearly, we see $|\zeta_\text{max}|$ goes to $1$ and back when entering and exiting the resonance.
As explained in the text, when $|\zeta_{\text{max}}|=1$ 
we have a dark state in the system, see equation~(\ref{eqdark}).
}
\label{fig:zm}
\end{figure}

We tested our theory using several model systems.
For example,
a benzene-type ring (equation (\ref{eq1}) with $L=6$),
as presented in Figure~\ref{fig:sixcomb}, 
where excellent agreement between the theory and numerically exact results is witnessed.
We see, as predicted by equations (\ref{eq11},\ref{eq12}), 
the width of the transition becomes smaller as the restart time $T_R$ grows.
%
To verify the uniqueness of $\zeta_\text{max}$,  
in Figure~\ref{fig:zm}, we present the behaviors 
of the eigenvalues $\{ \zeta_i \}$ for the model of benzene-type ring, 
when the sampling time $\tau$ is varied.
One of the eigenvalues, namely $\zeta_\text{max}$, approaches the unit circle when $\tau$ goes to $\pi/2$,
while the other pair of conjugate eigenvalues are relatively far from the unit circle.
As previously stated, when the largest eigenvalue $|\zeta_\text{max}|$ approaches the unit disk, 
the relevance of the other eigenvalues is negligible and the restart uncertainty relation presented in this work becomes relevant.

A natural query is to study the effects of system size on our main results. 
To this aim, we analyzed two models: 
the ring model and the complete graph with $L$ sites. 
The case $L=3$ corresponds to the experimental study we conducted. 
For $L > 3$, the results exhibit distinct behaviors. 
Focusing on the merging of two phases, 
corresponding to the largest and ground state energy, 
we find $w = 1+ L/2$ ($w=(1+L)/2$) for the even (odd) ring model 
and $w = 2$ for the complete graph.
Assuming the hopping amplitude $\gamma$ 
(as indicated in the inset of Fig. \ref{fig:Theomean} and equation (\ref{eq1})) 
is $L$-independent,
the width of the resonance decreases as we increase $L$ (see the SI).
However,
considering the resonance related to the first excited state and the ground state, 
for the ring model we find that the resonance width
will increase as the size of the system grows. 
The complete graph has merely two energy levels 
hence this choice of energy levels is clearly 
the same as the min-max choice, mentioned above.
The key issue for the broadening effect is how the energy gaps
and the parameter $\lambda$ scale with the size of the system. 
$w$ depends on the symmetry of the system and the degeneracy of the energy levels. 
For example, in the complete graph, the number of distinct energy 
levels is two for any $L$, which means $w = 2$. 
This results in relatively short mean hitting times in units of $\tau$ 
compared to the ring model. 
Importantly, these different behaviors are all captured 
by our time-energy-like restart uncertainty principle.

\section*{Effect of random perturbations}

The broadening of resonances in the first hitting time can arise from various sources. 
In the triangle model implemented on the IBM quantum computer, 
we have demonstrated that this broadening is attributable to the foundational principles of quantum theory and the restart paradigm. 
However, a broader objective is to explore the relationship 
between stochastic perturbations and these broadening effects,
and to determine whether the observed topological invariant $w$ 
is resilient to fluctuations of parameters. 
This investigation, whose details are provided
in the SI,  
encompasses fluctuating sampling times,
as well as randomness in restart times.

%
\begin{figure}[ht]
\begin{center}
\includegraphics[width=0.95\linewidth]{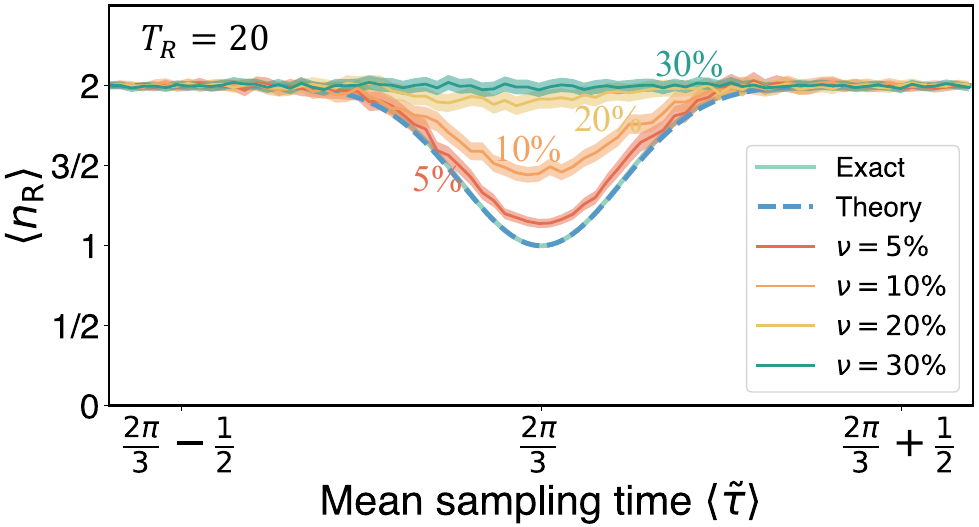}
\end{center}
\caption{ Mean hitting time versus the mean sampling time $\langle \tilde{\tau} \rangle $, for the three-site ring model 
with varying fluctuation levels in the evolution time \(\tau\) 
and fixed \(T_R = 20\). 
Utilizing the Monte Carlo method with \(30,000\) realizations, 
we find that as the fluctuations of $\tau$ increase, 
the resonances are progressively diminished, 
yet the topological number $\langle n_R \rangle = 2$, far from the resonance, remains unaffected and exhibits robustness.
}
\label{fig:randtau}
\end{figure}

Utilizing the three-site ring model, 
we studied the effect of random sampling time and random restart time on our key results. 
Using $T_R=20$, as we did in the experiment, 
allowing for fluctuations of up to five percent in the sampling time $\tau$
did not alter our main conclusions. 
However, when fluctuations in the sampling time $\tau$ reached $30$ percent, 
the dip in the resonances became difficult to observe, as shown in Figure \ref{fig:randtau}. 
There $\tilde{\tau}$ is the actual sampling time, uniformly distributed on $[\tau(1-\nu),\tau(1+\nu)]$, 
and $\nu$ indicates the fluctuation level. 
In addition, we found that the resonance is diminishing when $T_R$ is increased, 
for a fixed fluctuation level of $\tau$ (see Figure S15).
Thus, the larger $T_R$ is, the more pronounced the effects of random sampling times are.
Interestingly, the topological invariant far from the resonance, 
$\langle n_R \rangle \simeq w = 2$, 
remained robust even with significant fluctuations and large $T_R$, indicating the resilience of this number (see Figure \ref{fig:randtau} and Figure S15 in SI).
Similar behaviors are also observed for the benzene-type ring model, 
see Figure S14 in the SI.

To study the effects of random restart time $T_R$, we focused on two models, 
assuming $\langle T_R \rangle = 20$, motivated by our experiments. 
Using a narrow distribution of $T_R$ 
(a tent-like distribution) 
and a model where $T_R$ is Poisson distributed 
(a relatively wide distribution),
we show in SI that the effects of random $T_R$ are marginal (see Figures S16 and S17). 
This is because of two reasons: 
the location of resonances is insensitive to $T_R$,
as they are controlled by energies and the sampling time
and because we use (roughly) symmetric around the mean distributions for $T_R$.
It should be noted that the restart mechanism is a classical process, 
though one could extend it to consider a quantum coin-tossing process for the restart itself. 
In the SI, we outline the Pal-Reuveni framework \cite{Shlomi2017} for random and discrete restart times, suitable for our study.

Our findings show that the restart time-energy uncertainty relation does not change considerably
for the restart time distributed symmetrically about its mean, 
compared with the fixed restart time theory.
And this type of resilience also remains 
when the stroboscopicity of our measurement protocol is perturbed (fluctuating $\tau$) 
and when the measurement time $T_R$ is not vastly exceeding $40$ (for the fluctuation level $\nu=5\%$ which is already exaggerated on current-day quantum computers).
Notably, the topological number far from resonance
is robust to both significant fluctuations of $\tau$,
and long measurement time $T_R$.
Although the fluctuations in $T_R$ are not likely to happen in current quantum computing platforms,
we speculate that non-precise sampling times are not rare and might stem from noise and errors on quantum computers,
suggesting a wider range of applications of the restart uncertainty relation on noisy quantum simulation and computations.

\section*{Impact of quantum error and noise}

We now return to the issue of quantum error and noise existing in our experimental implementation.
Note that in our IBM experiments we used two qubits, see section Material and Methods.
This means that we have four states: 
$|01\rangle$, $|00\rangle$, $|10\rangle$, and $|11 \rangle$, 
where $|11 \rangle$ is theoretically decoupled from the other three 
while the first three states correspond to
the graph states $|0 \rangle,|1\rangle,|2 \rangle$, respectively.
By measuring the second qubit, 
we determined whether the system was in the target state $|01 \rangle$.
Ideally, the operations should isolate the system from $| 11 \rangle$, 
but noise existing on the quantum processors causes minor leakage into this state,
rendering the deviations in $P_{\rm det}$ as mentioned, and affecting the restart recurrence time. 
A key issue is to develop noise models that accurately capture the shift observed in Figure \ref{fig:Tridip}(b), 
necessitating a detailed analysis of the quantum circuit under consideration.
Incorporating IBM-provided noise models (see SI and \cite{ibmNM}), 
into the same quantum circuit employed in the experiment, 
namely a four-state model, 
we simulated this effect, revealing an upward shift in $\langle n_R \rangle$ (Figure S3 in SI).
This is consistent with our experimental findings (Figure \ref{fig:Tridip}(b)). 
More specifically, we incorporated bit-flip errors and thermal relaxation noise models (see SI).  
A key feature of these models, is the transfer
of amplitude to the theoretically forbidden state,
namely a leakage effect which is captured by the four-state model.

While the error in our experiment is roughly $1\%$, as mentioned, 
one might wonder what happens if the noise levels increase. 
We anticipate a transition of the recurrence time to its classical limit. 
The relevant classical theory, based on a random walk picture, 
suggests that for a two-qubit system like the one we used, with four states, 
we would expect $\langle n \rangle = 4$ 
according to Kac's theorem \cite{kac} when $T_R \to \infty$. In this classical limit, no resonances are observed. 
This discussion highlights that the quantum hitting times we measured are consistently shorter than this classical limit. 
Whether a quantum-to-classical transition in the first hitting times occurs 
due to increased noise levels remains an open question for future work.

\section*{Discussion}
In a broader perspective, 
the observed transitions exhibit similarities to 
line-shape resonances and broadening encountered across various fields of spectroscopy
\cite{cohen2019quantum1}. 
However, a distinguishing feature here is
that the periodic driving force is not an external field 
acting upon a material system.
Rather, they arise from the intrinsic nature of the measurements themselves 
and their periodicity.
Notably, resonances are associated with the creation of dark states, 
in contrast to traditional resonances linked to quanta of energy 
carried by particles such as photons.

Dark states are commonly observed in quantum systems, 
often appearing as dips in line shapes due to destructive interference, 
for example in electromagnetically induced transparency (EIT) 
\cite{harris1997electromagnetically,liu2001,Fleischhauer2005}
and coherent population trapping (CPT)
\cite{Alzetta1976,Whitley1976,Arimondo1976,gray1978coherent,Lounis1992,arimondo1996v} experiments. 
In the recurrence problems, where we measure the mean hitting time, 
these states play a unique role. 
Similar to the role of dark states in other fields, 
where they enhance effects like laser cooling \cite{Cohen1988,Cohen1995,Rio_2012}, 
the formation of dark states in our context leads to a speedup of the recurrence time.
This acceleration occurs because dark states
reduce the effective size of the Hilbert space, 
making searches more efficient and resulting in faster detection at resonances.
This holds true for the recurrence problem, 
namely the initial condition under study is detected with probability one 
if $T_R \to \infty$, so we are focusing on a bright state all along, 
though our observable $\expval{n}$ is clearly influenced
by the creation of dark states in the Hilbert space.

The broadening of the resonances of recurrence time is intricately linked to three crucial factors:
the uncertainty $\sigma_n$, 
the slowest decaying mode in the problem, i.e. $|\zeta_\text{max}|$, 
and the energies of a pair of merging phase factors.
This interconnection establishes fundamental relationships 
between quantum hitting time statistics and the system's underlying characteristics, 
with the restart time playing a pivotal role.
It is noteworthy that analogous resonances may be present in related scenarios,
particularly when we venture beyond the recurrence problem 
or engage in non-local measurements \cite{Didi2022}.
The expansion of our findings to encompass other observables 
and the exploration of cases 
where degeneracies are associated with the absolute value of the eigenvalue $|\zeta_\text{max}|$, 
resulting in non-pure exponential decay of $F_n$ 
and transitions from $w\to w-2$ or $w\to w-3$, etc., rather than the studied $w\to w-1$ case, 
represents an avenue for future research.

Additionally, we have devised a method for detecting resonances 
and quantifying their widths in the context of restarted hitting times on quantum computers. We anticipate this to be a valuable tool for investigating the interplay 
between mid-circuit measurements and unitary operations.
The width of the resonance can serve as an indicator of 
whether the fundamental postulates of measurement theory are effectively functioning 
on a given device or if noise and decoherence are exerting control.
In our experimental study, which was remotely conducted on an IBM quantum computer, 
we demonstrated that the former scenario holds true.
However, we anticipate that, as we increase the size of the quantum system 
or adjust the restart time, 
distinct behaviours related to the coupling of these systems to the environment may emerge.
Such insights will provide valuable information on the operating conditions 
of the new generation of algorithms with mid-circuit measurements, 
e.g. dynamic circuits \cite{bumer2023efficient} and error correction \cite{Krinner2022}.
Furthermore, quantum dynamics driven by measurements has emerged as an intriguing method 
to study novel phenomena, for example, entanglement transitions \cite{PhysRevX.9.031009,Koh2023}, 
induced chirality \cite{PhysRevB.108.214305}, 
and synchronization \cite{PhysRevLett.132.010402}.
When implemented on a quantum computer, 
finite-time effects and hence restart will likely emerge as important.

The strategy of restarts used here is nearly mandatory for several reasons.
In real quantum circuits, noise and leakage are present.
Hence to study the quantumness of the problem, one is obliged to use finite-time experiments.
More generally, unless one finds a way to perfectly correct noise and eliminate leakage 
in quantum computers with mid-circuit measurements, 
the restart strategy is nearly a must.
The significance of the broadening effect becomes crucial 
close to discontinuous behaviours of the hitting time statistics, 
leading to a time-energy uncertainty relation deeply related to 
the variance of the first detection time.
This insight, promisingly, holds the potential to contribute to 
a better understanding and design of efficient quantum algorithms, 
which rely on backtracking (restart) and monitored dynamics \cite{Montanaro2018}.
More importantly, 
we provided a restart hitting time uncertainty relation, 
and since hitting times are fluctuating, 
the uncertainty relation differs from the standard time-energy relation,
where time is a parameter and not an observable.


\section*{Materials and Methods}
\subsection*{Model}
The example we considered in the main text is a ring model 
governed by the nearest-neighbor tight-binding Hamiltonian
\begin{equation}
\label{eq1}
H = -\gamma \sum_{j=0}^{L-1} \left( \ket{x} \bra{x+1}+ \ket{x} \bra{x-1} \right),
\end{equation}
where $\gamma$ is the hopping amplitude, $L$ is the size of the system,
and $\{\ket{x}\}$ are the spatial states composing the ring system.
As noted, the main results in the manuscript are generally valid and are not limited to this model.  
The periodical boundary condition indicates $\ket{0} = \ket{L}$, and $\ket{0}$ is the target state.
The eigenvalues of the Hamiltonian (\ref{eq1}) are
\begin{equation}
\label{eigenvalues1}
\begin{aligned}
    E_k = - 2 \gamma \cos \theta_k, 
\end{aligned}
\end{equation}
with $\theta_k = {2 \pi k / L}$ and $k=0,1,2,\dots, L-1$.
The corresponding eigenstates are 
$\ket{E_k} = \sum_{x=0}^{L-1}e^{i\theta_k x} \ket{x}/\sqrt{L}$.
Hence the overlap is $\left| \braket{x}{E_k}\right|^2 = 1/L$.
In the main text, for simplicity, we set the hopping amplitude $\gamma$ as $1$.

{\em The three-site ring}
was used in our remote IBM experiments.
Using equation (\ref{eigenvalues1}) with $L=3$,
there are $2$ distinct energy levels, $\{ -2, 1 \}$,
with $|\langle x|E_k\rangle|^2=1/3$ and the energy level $E_1=E_2=1$ is doubly degenerate.
Hence the overlaps are $p_{-} = 2/3$, and $p_{+} = 1/3$.
When $\tau = 2  \pi j /3$ with $j = 0, 1, 2,\dots$, 
the mean $\expval{n}$ for $T_R \to \infty$
jumps from $w=2$ to $w=1$, 
where energy phases $\{e^{-i \tau}, e^{i 2 \tau}\}$ match.
Using the above mentioned $p_-$ and $p_+$ and energies, 
equations (\ref{eq11},\ref{eq12}) give $\lambda = 2/9$, 
and $\widetilde{\Delta E \tau } = \left| 3\tau-2\pi j \right|$ 
close to each $\tau=2  \pi j /3$.
In Figure~\ref{fig:Tridip}, $j=1$ and the resonance condition $\tau = 2\pi/3$ is used.
As mentioned, these jumps in the mean hitting time correspond to revivals of the wave packet on the origin.

{\em The benzene-type ring}
was used in our examples plotted in Figure \ref{fig:sixcomb}.
Here $L=6$ and 
%
the distinct energies are 
$\{\pm 2, \pm 1\}$ where the energies $\pm 1$ are two-fold degenerate.
Hence the overlaps corresponding to distinct energies are 
$|\braket{0}{E_0=-2}|^2 = |\braket{0}{E_3=2}|^2 = 1/6$, 
and $|\braket{0}{E_1=-1}|^2 + |\braket{0}{E_5=-1}|^2  =
|\braket{0}{E_2=1}|^2 + |\braket{0}{E_4=1}|^2 = 1/3$. 
Using equation (\ref{eq01}) we therefore expect that,
except for a small subset of $\tau$'s, $\expval{n} = 4$.
When $\tau={(2j+1)\pi/ 2}$ with $j=0,1,2,\dots$,
$\expval{n}$ for $T_R \to \infty$ jumps from $w=4$ to $w=3$, 
where the energy phases $\{e^{i2\tau},e^{-i2\tau}\}$ merge,
hence $E_+$ and $E_-$ used in the text are $-2$ and $2$, respectively. 
So the parameters in equations (\ref{eq11}, \ref{eq12}) are,
$\lambda = 3/4$, 
and $\widetilde{\Delta E \tau } = \left| 4\tau - 2\pi (2j+1) \right|$ 
close to each $\tau= (2j+1)\pi/ 2$.
In Figure~\ref{fig:sixcomb}, $j=0$ or $\tau = \pi/2$ is used.

\begin{figure}[t]
\begin{center}
\includegraphics[width=0.9\linewidth]{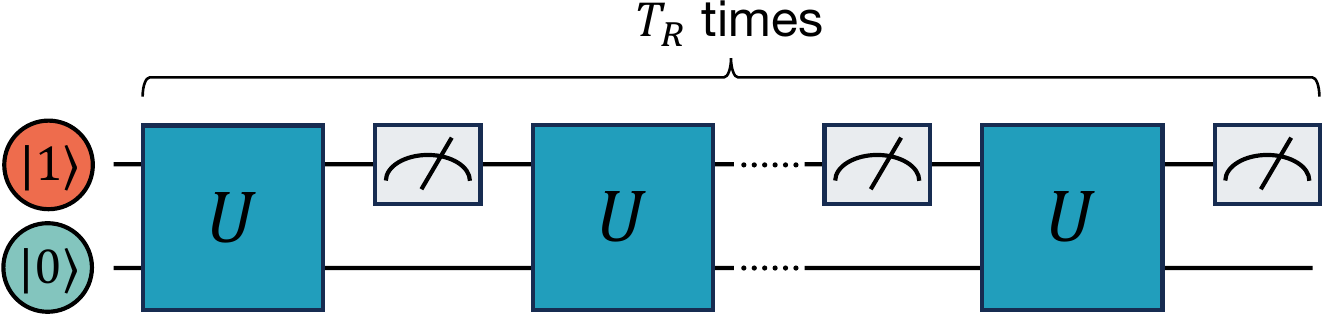}
\end{center}
\caption{{Quantum circuit representation for the three-site ring model.} 
Quantum circuit for two qubits representing the three localized states 
with alternating unitary $U$ and measurements, 
with the initial state and target state $\ket{0} = \ket{01}$.
}
\label{fig:circuit}
\end{figure}

\subsection*{Sketch of the rigorous proof for the uncertainty relation}
To prove the uncertainty relation, 
the key is to validate equation (\ref{eq07}).
Briefly speaking, this can be done via the generating function method \cite{Friedman2017a}.
Applying the $Z$-transform to the expression inside the bracket of equation (\ref{eq01a}), 
i.e. $\phi_n = \bra{0} U(\tau){\cal S}^{n-1} \ket{0}$,
one can obtain the generating function, $\tilde{\phi}(z) = \sum_{n=1}^\infty z^n \phi_n$.
Decomposed by the Hamiltonian's eigenstates, and being a polynomial, 
$\tilde{\phi}(z)$ can be factorized 
by its zeros
and poles, 
using Blaschke factorization \cite{Gruenbaum2013}.
Due to the mathematical property of the latter, 
the poles are the reflection of the zeros, with respect to the unit circle.
And also, the zeros
are the complex conjugate of the eigenvalues, 
$\{\zeta_i\}$, 
of the survival operator ${\cal S}$ (see Supplementary Note 2 in SI) \cite{Thiel2020D}.
Hence, the generating function $\tilde{\phi}(z)$ can be completely factorized 
by the zeros, or the eigenvalues $\{\zeta_i\}$.
This allows us, in terms of $\{\zeta_i\}$, to use the residue theorem, 
to recover $\phi_n$ via the inversion formula 
$\phi_n = \frac{1}{2 \pi i}\oint_{|z|=1} \tilde{\phi}(z)\, z^{-(n+1)}\, dz$.
And then $F_n=\left| \phi_n \right|^2$ can be computed and simplified to equation (\ref{eq07}). 
The detailed derivation 
is presented in the Supplementary Note 2 in SI.

\subsection*{Implementation on a quantum computer}

We design a three-site ring model, Figure \ref{fig:Theomean}, 
using equation (\ref{eq1}) with $L=3$.
To realise the three-site system on a quantum computer, 
we use two qubits, which can generate four states: 
$\ket{00}, \ket{01}, \ket{10}$ and $\ket{11}$. 
Hence, we employ the following mapping between the qubits and spatial states representation:
$\ket{01} \rightarrow \ket{0}, \ket{00} \rightarrow \ket{1} \ \mbox{and} \
\ket{10} \to \ket{2}$.
We design our circuit in such a way that the additional state $\ket{11}$ is not connected to the others 
and will never be detected at least theoretically. 

In our study we detect the state $\ket{0} \to \ket{01}$.  
This can be realised by measuring only the upper (right) qubit.
Hence, when measuring the upper (right) qubit in state $\ket{0}$, 
the measurement does not give any information to distinguish 
the state $\ket{1} \to \ket{10}$ and $\ket{2} \to \ket{00}$. 
Importantly, measuring the upper (right) qubit in state $\ket{1}$ tells that
the system is in $\ket{0} \to \ket{01} $ with certainty.

We determine the first detection time, $n$, by analysing mid-circuit measurement outputs from the quantum circuit, as shown in Figure~\ref{fig:circuit}. 
We examine the expected value of $n$ as a function of $\tau$,
considering the detection of the target state, 
namely the upper (right) qubit being detected in state $\ket{1}$, 
as the endpoint of measurement.
As detailed earlier, measurements restart at finite \(T_R\), 
yielding output strings like $\{0, 1, 0, 1, 1, \dots\}$, of length $T_R$,
with ``$0$'' and ``$1$'' indicating the state of the upper (right) qubit, 
or actually failure and success in detection, respectively. 
The experiment ideally concludes after the first appearance of ``$1$'', 
but due to technological constraints, 
we cannot terminate the quantum computation based on the measurement outputs, 
necessitating a finite and constant \(T_R\).

The maximum duration for measurement repetitions in the IBM quantum computer 
IBM Sherbrooke is set at \(T_R \simeq 20\) . 
This restriction is influenced by software limitations specific to the quantum computer we used. 
This choice is also chosen to reduce noise and avert non-unitary actions 
and probability leakage.
Such occurrences could render the system's Hamiltonian ($H$) effectively non-Hermitian. 
In particular, when performing our experiments on IBM Sherbrooke, 
$T_R = 20$ was the maximum number of repeated measurements allowed by the software.

As shown in Figure~\ref{fig:protocol}, 
to calculate the conditional mean $\expval{n}_{\text{Con}}$, 
we disregarded null-detection strings, 
which are strings of length twenty with only zeros $\{ 0, 0, \dots, 0\}$. 
Such strings are rare, since the $P_\text{det}$ within $20$ measurements is nearly $1$ 
(at most $2$ percent below $1$, depending on $\tau$), 
see details and figure for $P_\text{det}$ in SI.
For the restarted mean, we analyse the unconditional hitting time with restarts,
noting the first detection time as $n_R$. 
For example, consider the sequence of $\{0, 0, \dots, 0 \}$ of length $20$,
which, after a restart event, is followed by $\{ 0, 0, 1, \dots \}$.
Here, the first time for detection under restart is $n_R = 23$. 
In total, we conducted $32,000$ runs with $T_R = 20$ bits per run,
requiring additional data processing to identify the first ``$1$'' in each string, 
thus obtaining the first hitting time $n$ for each run.
See the Supplementary Note 5 in SI for more details on the quantum circuit implementation,
error suppression, and data processing.

\subsection*{Data, Materials, and Software Availability}
The experimental data generated in this study are available 
at
\url{https://doi.org/10.5281/zenodo.13327746}.

%
\begin{acknowledgments}
We acknowledge the use of IBM Quantum services. 
The views expressed in this work are those of the authors
and do not reflect the official policy or position of IBM or the IBM Quantum team.
Q.W. and R.Y. acknowledge the use of the IBM Quantum Experience and the IBMQ-research program. 
Q.W. would like to thank the Max Planck Institute for the Physics of Complex Systems for its hospitality. 
The support of Israel Science Foundation’s grant 1614/21 is acknowledged.
\end{acknowledgments}

\bibliography{ref}

\clearpage

\onecolumngrid
\appendix 

\newcommand{\supplementarytitle}{
    \begin{center}
        \Large\bfseries Supplementary Information for \\ 
``Restart uncertainty relation for monitored quantum dynamics''
    \end{center}
    \vspace{0.10cm} 
}

\supplementarytitle

\section*{1. Experimental mean hitting time under restart and noise model simulations}

We now address the origin of the vertical shift observed 
in the experimentally derived restarted mean hitting time, $\expval{n_R}$, 
as presented in Fig.~3(b) in the main text. 
Given the observed strong concordance between experimental results and exact calculations of $\expval{n}_\text{Con}$, as shown in Fig. 3(a), 
we postulate that the vertical shift primarily stems from the second term in equation (4), 
$T_R[1-P_\text{det}(T_R)]/P_\text{det}(T_R)$, 
in the context of quantum hitting time with restarts.

To substantiate this hypothesis, 
we illustrate the detection probability $P_\text{det}(T_R)$ with $T_R=20$ 
in Fig. \ref{fig:nshiftpdet}, 
obtained from experiments, exact calculation (using equation (2)), 
the theory (see below) and simulations (using IBM quantum simulators).
Using equation (8) in the main text,
and $a(\zeta_{\text{max}}) = ( 1- | \zeta_{\text{max}} |^2 )^2$, 
we get the theory
\begin{equation}\label{pdet}
    P_\text{det}(T_R) = \sum_{n=1}^{T_R}F_n 
    \simeq 
    1 - ( 1 - \left| \zeta_{\text{max}} \right|^2 ) 
    e^{-T_R (1- \left| \zeta_{\text{max}} \right|^2)},
\end{equation}
The figure reveals a small discrepancy 
between the experimental and exact/theoretical/simulated results,
suggesting the presence of measurement noise.
More specifically, consider $ \tau$ far from resonance at $ \tau=2 \pi/3$,
the theory predicts $P_\text{det} \to 1$, 
namely within $20$ measurements the click yes is nearly surely guaranteed. 
The result from the experiment is $P_\text{det} \simeq 0.99$, 
namely the deviation from theory is merely one percent. 
However, using $P_\text{det}=0.99$ we get for $T_R=20$, 
$T_R(1-P_\text{det})/P_\text{det} \simeq 0.2$, while the theory predicts a nearly zero value.
This means that $\expval{n_R}$ is expected to be shifted 
by roughly $0.2$ due to the small error in $P_\text{det}$. 
The issue here is that a small variation in $P_\text{det}$, or the order of one percent,  
can lead to a small shift for the mean return time, 
since the second term in equation~(4) is linear in $T_R$. 
The larger $T_R$ is the bigger we expect the shift in $\expval{n_R}$ to be.
Remarkably, the shift of $P_{\text{det}}$ is roughly one percent for all $\tau$, 
see Fig. \ref{fig:nshiftpdet}. 
It follows that the shift of $\expval{n_R}$ due to the small errors in $P_\text{det}$, 
is roughly $0.2$. 
To test this we plot in Fig. \ref{fig:shiftrestart}, $\expval{n_R}-\text{shift}$, 
where as mentioned for $T_R=20$, the expected shift is $0.2$. 
Now the theory and exact results reach an excellent agreement with the experimental results.

It is crucial to highlight that in the data analysis of the conditional mean $\expval{n}_\text{Con}$, as described by equation (3), the noise in $P_\text{det}(T_R)$ is effectively mitigated or eliminated through the exclusion of non-detection trajectories, as $P_\text{det}(T_R) = \sum_{n=1}^{T_R} F_n$ appears in the denominator of the equation This further explains the observed perfect alignment between the theoretical prediction and experimental results for $\expval{n}_\text{Con}$.

\begin{figure}[h]
\centering
\includegraphics[width=0.45\textwidth]{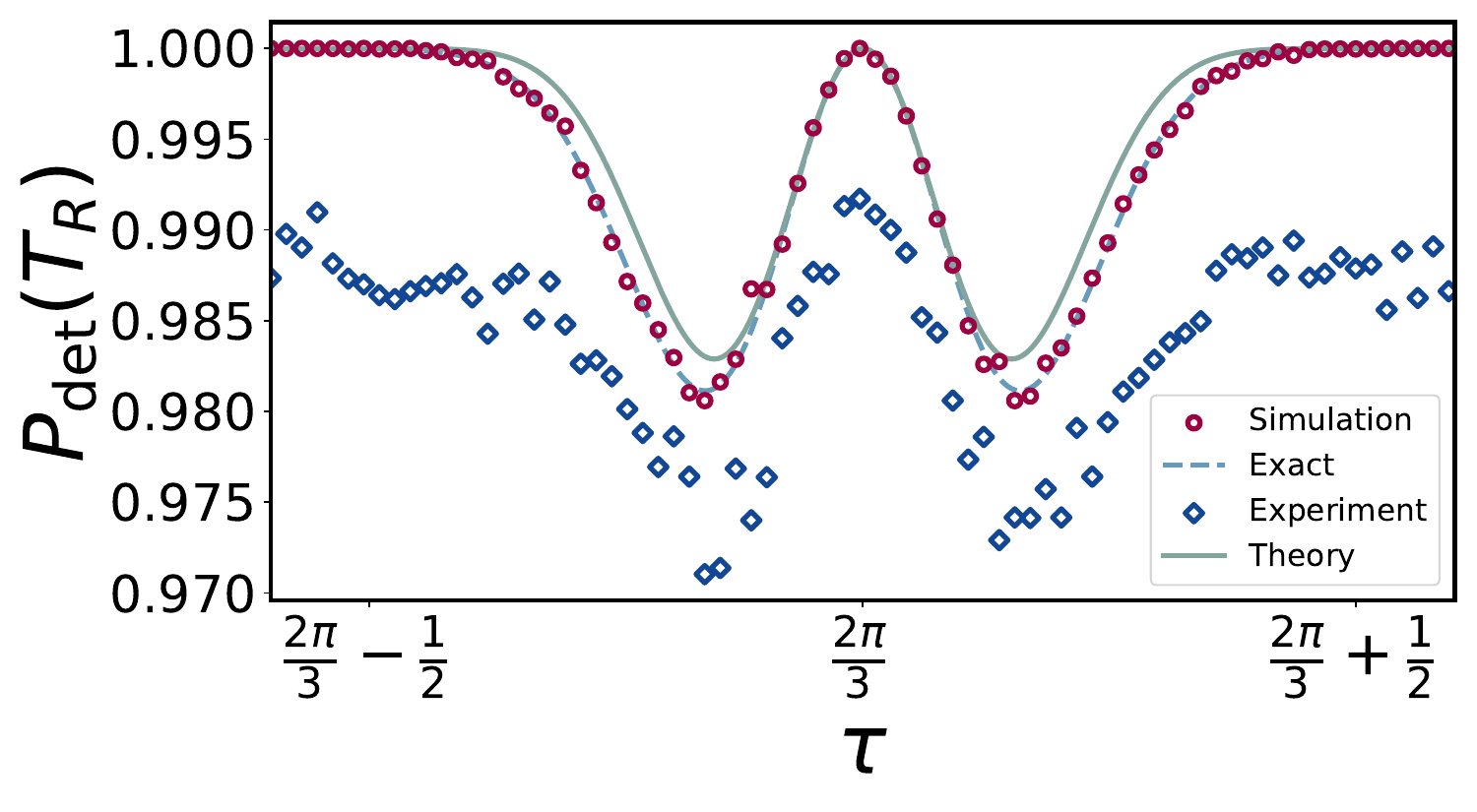}
\caption{
The sample $P_\text{det}(T_R) = \sum_{n=1}^{T_R} F_n$, for $T_R = 20$, 
for quantum hitting times under restarts versus $\tau$ is estimated from the experimental data.
The red circles are obtained from IBM quantum simulators, 
and the blue dots are from experiments on the IBM quantum computer. 
The green solid/dashed line represents 
the theory equation (\ref{pdet})/exact results (using equation (2)).
We see that experimental results are shifted compared to theory,
revealing roughly one percent error in the measurement. 
This error gives rise to the shift observed in Fig. \ref{fig:shiftrestart}.
The model here is a tight-binding three-site ring as in Fig. \ref{fig:shiftrestart}.
}
\label{fig:nshiftpdet}
\end{figure}
\begin{figure}[h]
\centering
\includegraphics[width=0.95\textwidth]{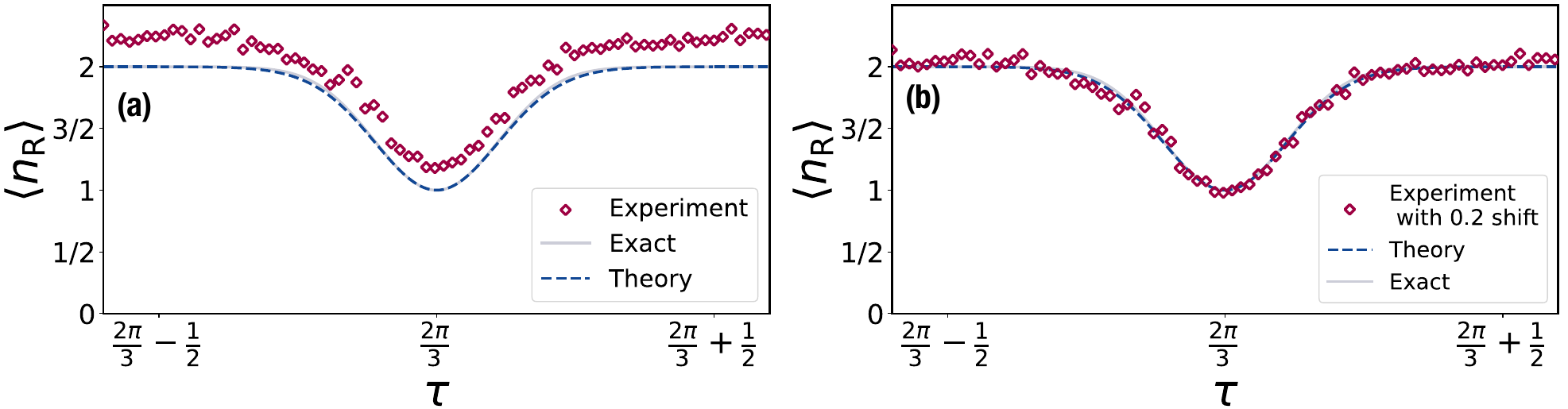}
\caption{(a) The mean hitting time, in units of $\tau$, under restart, $\expval{n_R}$, as a function of $ \tau$. 
We compare the exact results (obtained by equation (4), gray solid line), 
the theory (obtained by equation (7), blue dashed line) 
and experimental results on a quantum computer (squares) for $T_R = 20$.
We observe the vertical shift between the experimental and exact results, 
which is due to noise in the quantum computer.
The model here, as in Figs.~2,3 in the main text, 
is a tight-binding three-site ring (equation (15) with $L=3$). 
(b) The down-shifted experimental $\expval{n_R}$ (squares) compared with the theory (the dashed line obtained from equation (7)) and exact results (the solid line obtained with equation (4)).
With the experimental data shifting downward by $0.20$ explained by Fig. \ref{fig:nshiftpdet}, the theory agrees nicely with the experimental results.
}
\label{fig:shiftrestart}
\end{figure}

We now analyze the cause of errors in $P_{\rm det}$, which is related to quantum error and noise and its consequent leakage.

{\bf Leakage.} 
In the implementation of the three-site ring model using mid-circuit measurements 
on the IBM quantum computer, we employed a two-qubit system. 
As mentioned in the main text, in our model, 
the states of the triangle model, $\ket{0}$, $\ket{1}$ and $\ket{2}$, 
are mapped to the qubit states $\ket{01}$, $\ket{10}$ and $\ket{00}$, respectively. 
Theoretically, the state $\ket{11}$ is decoupled from the other states. 
However, practical experiments on a quantum computer demonstrated 
leakage from the utilized qubit states ($\ket{00}, \ket{01}, \ket{10}$) 
to the excluded state ($\ket{11}$), 
as mentioned in Materials and Methods. 
Note that after twenty measurements (which is the length of our experiment),  
we find leakage of one percent, hence while clearly an important issue, 
the leakage is not large.
We anticipate an increase of leakage as we increase $T_R$ and possibly also if the size of the system grows
as more noise will be present. 
This in turn will affect the mean recurrence  time.
We want to note
that the leakage in our problem is merely one of the consequences of noise 
existing on current quantum processors. 

\begin{figure}[ht]
    \begin{center}
    \includegraphics[width=0.95\linewidth]{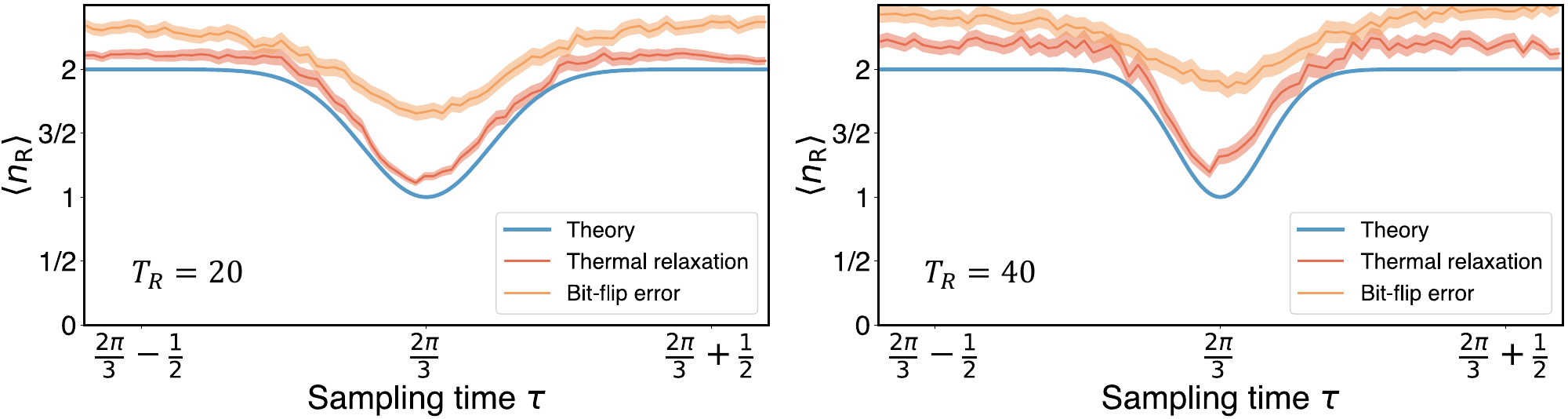}
    \end{center}
    \caption{
    The effects of noise on the mean recurrence time for the three-site ring model.
    We chose two common noise models provided by the IBM quantum computing platform,
    namely the bit-flip error and thermal relaxation noise models \cite{ibmNM}, 
    with parameter values chosen to align with the IBM technical document 
    (see Ref. \cite{ibmNM} for technical details).
    The blue curve represents the theoretical $\expval{n_R}$ with no noise. 
    A vertical shift is witnessed for the thermal relaxation noise, 
    while an additional diminishing resonance is presented for the bit-flip error.
    These noise-induced effects are more pronounced for a longer restart time,
    since the noise accumulates with measurement time.
    The results are obtained using IBM simulators. 
    For bit-flip error, we choose ``strong'', and for thermal relaxation, we choose ``moderate'' noise levels (see parameters in Table \ref{tab:noise}).
    }
    \label{fig:noisemodel}
\end{figure}

\textbf{Noise on IBM quantum processors.}
Noise in current quantum computing platforms is a critical challenge 
impacting computational accuracy and reliability. 
Quantum noise arises from various sources, 
including environmental decoherence, control errors, 
and imperfect quantum gate operations.
As mentioned, the leakage in our problem 
is merely one of the consequences of noise existing on current quantum processors. 
Fortunately, the IBM quantum computing platform provides various noise models,
with which one may predict behaviors of simulated quantum dynamics,
on noisy quantum circuits \cite{ibmNM}.
Utilizing two common noise models, 
i.e. bit-flip error, and thermal relaxation \cite{ibmNM},
we observed clear effects of noise on the mean recurrence time 
for the monitored quantum dynamics.
As seen in Figure \ref{fig:noisemodel},
an upward shift of the theoretical $\expval{n_R}$ 
is induced by these noise models,
yet the resonance dip remains visible.
Both noise accumulates with measurements and evolution time, 
resulting in more pronounced effects as $T_R$ increases, i.e. compare shifts on the left and right panels in Figure \ref{fig:noisemodel}.
It is noteworthy that our simulation is based on the same quantum circuit which is employed to conduct the IBM experiment in the main text, 
namely we have two qubits and hence four states that evolve on the noisy circuit. 
Therefore, we believe that our simulation is a proper estimator 
for a noisy quantum computing platform.

Now we present details of the noise models, 
including their physical implication, parameters and additional numerical results.
As mentioned above, we chose two noise models: 
the bit-flip error, 
and the thermal relaxation \cite{ibmNM}.

The bit-flip error noise model represents 
a quantum error that
probabilistically flips a qubit state, 
i.e. from $\ket{0}$ to $\ket{1}$ or vice versa, 
serving as a fundamental noise channel that explains state transitions.
This error might be led by gate imperfections, interactions with nearby qubits, etc.
The bit-flip error noise model is characterized by the following parameters and we extract from \cite{ibmNM} the description:

\begin{itemize}
    \item For a single-qubit gate, invert the qubit's state with a probability of \( p_{\text{gate1}} \).
    \item For a two-qubit gate, introduce single-qubit errors independently to each qubit.
    \item When resetting a qubit, set it to 1 instead of 0 with a probability of \( p_{\text{reset}} \).
    \item During a qubit measurement, flip the qubit's state with a probability of \( p_{\text{meas}} \).
\end{itemize}

We note that this model captures errors caused by measurements, 
as indicated by the parameter \( p_{\text{meas}} \), which is in line with our setup of repeated measurements.

The thermal relaxation noise model describes how a qubit state naturally decays over time 
due to interactions with its environment. 
This model encompasses two primary processes, 
energy relaxation or amplitude damping (also called \(T_1\) relaxation), 
and dephasing (or \(T_2\) relaxation).
The physical meaning of the parameters $T_1$, $T_2$ is the following:

\begin{itemize}
    \item $T_1$ relaxation is the process by which a qubit in the excited state $\ket{1}$ decays to the ground state $\ket{0}$. This represents the loss of energy from the qubit to the environment. 
    Over time, the probability of the qubit being in $\ket{1}$ decreases, leading to a loss of coherence in quantum computations.
    \item $T_2$ relaxation process describes the loss of phase information without a change in the energy level of the qubit, 
    e.g. the relative phase between $\ket{0}$ and $\ket{1}$ may change unpredictably, leading to decoherence.
    \item Longer $T_1$, $T_2$ times imply that the qubits can maintain their quantum state for longer periods, namely higher fidelity.
\end{itemize}
\noindent See the implementation of the two noise models using {\em Qiskit} in \cite{ibmNM}.

\begin{figure}[ht]
    \begin{center}
    \includegraphics[width=0.45\linewidth]{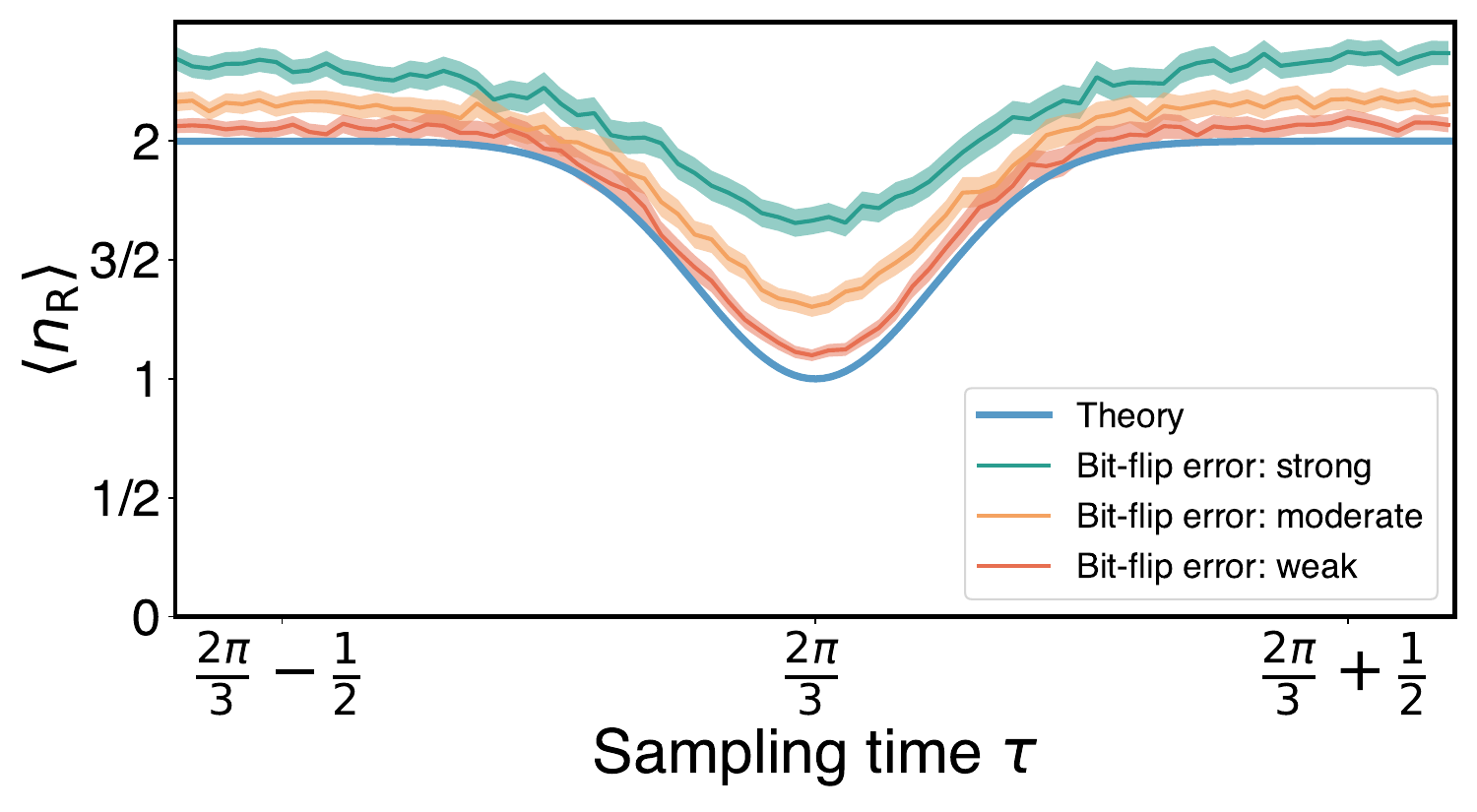}
    \end{center}
    \caption{The mean recurrence time exhibits diminishing resonance 
    when the circuit implementation is incorporated with bit-flip errors.
    Here the restart time $T_R=20$.
    We see that stronger noise leads to more pronounced effects,
    but the resonance, as well as the constant mean recurrence time far from the resonance,
    are not ruined by noise.
    We also see a shift upwards, compared to theory, as explained in the text. 
    See Table \ref{tab:noise} for parameters corresponding to ``strong'', ``moderate'' and ``weak'' noise. 
    }
    \label{fig:bitflip}
    \end{figure}

\begin{figure}[ht]
    \begin{center}
    \includegraphics[width=0.45\linewidth]{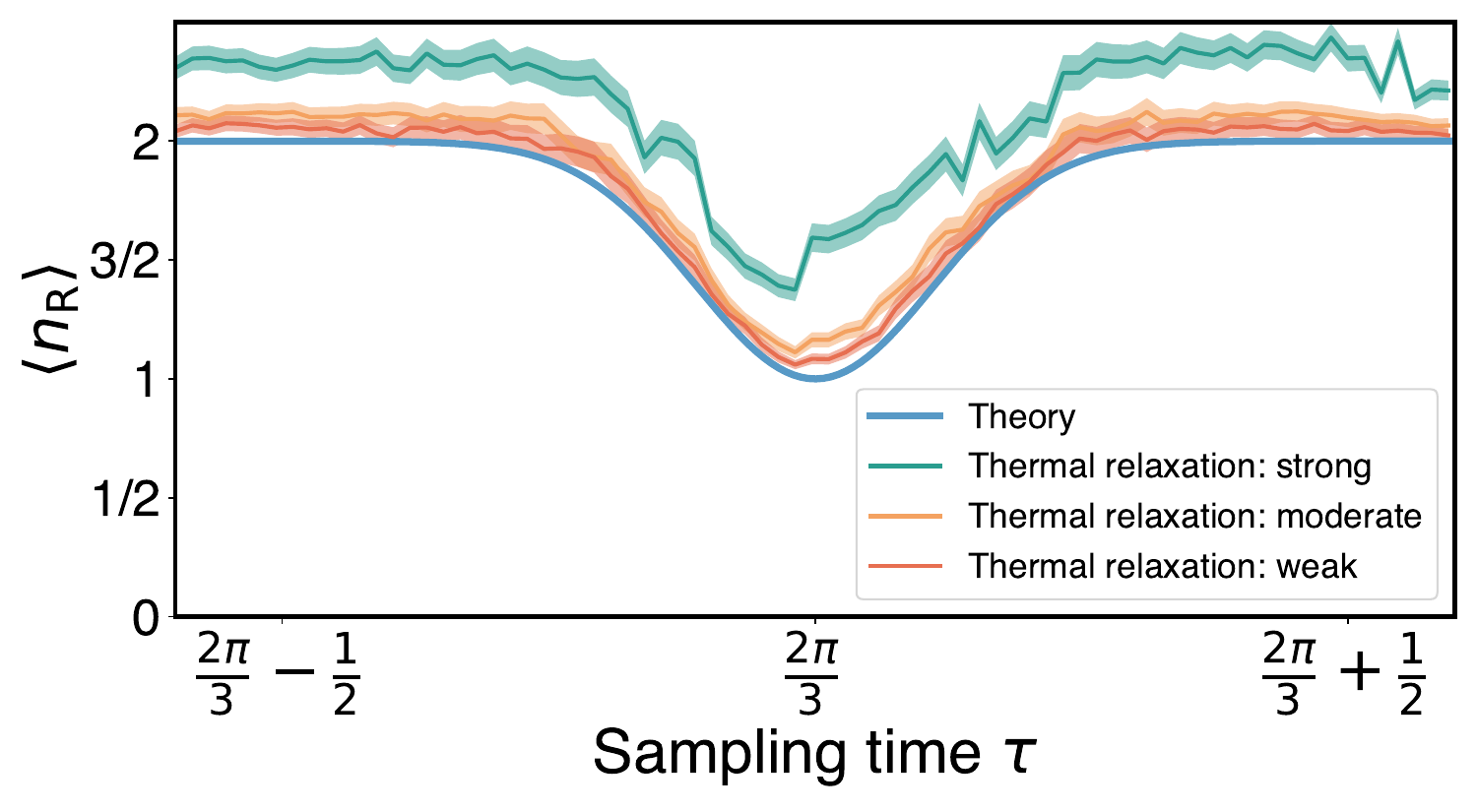}
    \end{center}
    \caption{The mean recurrence time is shifted vertically 
    when the circuit implementation is incorporated with the thermal relaxation model.
    Here the restart time $T_R=20$.
    The resonance,
    as well as the constant $\expval{n_R}$ far from the resonance,
    are relatively robust to this type of noise,
    except for an upward shift increasing with stronger noise.
    See Table \ref{tab:noise} for parameters corresponding to ``strong'', ``moderate'' and ``weak'' noise. 
    }
    \label{fig:therm}
    \end{figure}

For each noise model, 
we choose three set of parameter values, 
denoted as ``strong'',``moderate'' and``weak''
according to the noise strength, 
as specified in Table \ref{tab:noise}.
With these choices of parameters, we present in 
Figures \ref{fig:bitflip},\ref{fig:therm} 
the corresponding behaviors of the mean recurrence time.
As expected, 
intensifying noise leads to more pronounced results, 
e.g. larger upward shift and increasingly diminishing resonance.
For the bit-flip error, both shift and diminishing resonance are witnessed,
but the resonance dip remains visible,
while for the thermal relaxation noise, we mainly find the upward shift. 


\begin{table*}[t]
    \centering
    \caption{Parameters for the noise models used in Figures \ref{fig:bitflip},\ref{fig:therm}.}
\label{tab:noise}
\sisetup{detect-weight,mode=text}
\renewrobustcmd{\bfseries}{\fontseries{b}\selectfont}
\renewrobustcmd{\boldmath}{}
\newrobustcmd{\B}{\bfseries}
\begin{tabular}{ cccccccccccccccccccc } 
    \Xhline{2\arrayrulewidth}
    \multicolumn{4}{c}{ \B \,\,\,Noise models\,\,\,} & 
    \multicolumn{4}{c}{\makecell{\B Bit-flip error}} & 
    \multicolumn{4}{c}{\makecell{\B Thermal relaxation}} \\~\\
    \multicolumn{4}{c}{``Strong''} & 
    \multicolumn{4}{c}{$p_{\text{gate1}} = 0.005$, $p_{\text{reset}} = 0.003$, $p_{\text{meas}} = 0.01$}  & 
    \multicolumn{4}{c}{$T_1(\text{microsec}) \in [100, 20]$, $T_2(\text{microsec}) \in [140, 20] $}   \\~\\
    \multicolumn{4}{c}{``Moderate''} & 
    \multicolumn{4}{c}{$p_{\text{gate1}} = 0.001$, $p_{\text{reset}} = 0.001$, $p_{\text{meas}} = 0.01$} & 
    \multicolumn{4}{c}{$T_1(\text{microsec}) \in [250, 50]$, $T_2(\text{microsec}) \in [350, 50] $} \\~\\
    %
    \multicolumn{4}{c}{``Weak''} & 
    \multicolumn{4}{c}{$p_{\text{gate1}} = 0.0005$, $p_{\text{reset}} = 0.0005$, $p_{\text{meas}} = 0.001$} & 
    \multicolumn{4}{c}{$T_1(\text{microsec}) \in [500, 100]$, $T_2(\text{microsec}) \in [700, 100] $} \\
    %
    \Xhline{2\arrayrulewidth}
   \end{tabular}
\end{table*}

%

\newpage

\section*{2. Rigorous proof of uncertainty principle}

We will provide a rigorous proof for the above uncertainty relations, equations (6,7) in the main text.
To do so we will find $F_n$ in the large $n$ limit. We also find an exact expression for $F_n$.
%
In the following derivation, we note that equations~(\ref{renew}-\ref{neqw}) are not new.
The expression inside the bracket in equation (2) can be rewritten as \cite{Friedman2017a},
%
\begin{equation}
    \phi_n = \bra{0}\hat{U}(n \tau)\ket{0}
    -
    \sum_{m=1}^{n-1}\bra{0}\hat{U}((n-m) \tau)\ket{0} \phi_m.
    \label{renew}
\end{equation}
Here $\phi_n$ is the first detection amplitude,
and $F_n=\abs{\phi_n}^2$.
equation (\ref{renew}) is also called the quantum renewal equation
$\ket{0}$ is the initial and also the target state of the quantum walker. 
In our examples, the target state is a node on the graph, 
and since we have in these examples translational invariance, any node will hold.
%
%
Since equation~(\ref{renew}) has a convolution term,
applying the $Z$ transform, namely,
\begin{equation}
    \Tilde{\phi}(z) := \sum_{z=1}^{\infty}z^n \phi_n,
    \label{ztrans}
\end{equation}
we obtain the generating function \cite{Friedman2017a}
\begin{equation}
    \Tilde{\phi}(z) = \frac{\bra{0}\hat{\mathcal{U}}(z)\ket{0}}{1 + \bra{0}\hat{\mathcal{U}}(z)\ket{0}},
    \label{generfunc}
\end{equation}
where $\hat{\mathcal{U}}(z) := \sum_{n=1}^{\infty}z^n \hat{U}(n \tau) 
= z e ^{-iH \tau}/(1-ze^{-iH\tau}) $.
The generating function is a useful tool with which we may obtain many results, 
the inversion formula 
 \begin{equation}
     \phi_n = \frac{1}{2 \pi i}\oint_{|z|=1} \frac{dz}{z^{n+1}}\Tilde{\phi}(z)
     \label{inversion}
 \end{equation}
provides a formal solution to the problem. 
Via spectral decomposition of equation (\ref{generfunc}) (into the energy eigenbasis), 
we have \cite{Friedman2017a}
\begin{equation}\label{spec}
    \tilde{\phi}(z) = 
    {\sum_{k=1}^w \sum_{l=1}^{g_k} |\bra{0}\ket{E_{kl}}|^2
    {ze^{-iE_k\tau}/(1-ze^{-iE_k\tau})} 
    \over 
    \sum_{k=1}^w \sum_{l=1}^{g_k} |\bra{0}\ket{E_{kl}}|^2 (1-ze^{-iE_k\tau})^{-1}},
\end{equation} 
where $w$ is 
the number of distinct energy phase factors $\exp(-iE_k\tau)$ 
with non-zero overlap $\sum_{l=1}^{g_k} |\bra{0}\ket{E_{kl}}|^2$, 
$g_{k}$ is the degeneracy of $E_{k}$ ($g_k\ge2$ means degenerate energy levels),
and $\ket{E_{kl}}$ are the eigenstates corresponding to $E_k$.
equation (\ref{spec}) can be rewritten as
\begin{equation}\label{phiND}
\begin{aligned}
    \tilde{\phi}(z) &= {{\cal N}(z) \over {\cal D}(z)}, \\
    \text{with } 
    {\cal N} (z) 
    &=  z \sum_{k=1}^w \sum_l^{g_k} \left| \bra{0}\ket{E_{kl}} \right|^2
        \prod^w_{j=1,j\neq k} \left( z - e^{iE_j\tau} \right), \\
    {\cal D} (z)
    &=  \sum_{k=1}^w e^{iE_k\tau} \sum_l^{g_k} 
        \left| \bra{0}\ket{E_{kl}} \right|^2
        \prod^w_{j=1,j\neq k} \left( z - e^{iE_j\tau} \right).
\end{aligned}
\end{equation}
%
%
And one can prove the relation \cite{Friedman2017a}
%
\begin{equation}
    {\cal D}(z) = (-1)^{w-1} e^{i\chi} z^w \left[ {\cal N} \left( 1/z^* \right) \right]^*,
\end{equation}
where $\chi = \sum_{k=1}^w \tau E_k$, and the superscript ``$\ast$'' means complex conjugate.
Then we can factorize $\tilde{\phi}(z)$ as \cite{Gruenbaum2013}
%
\begin{equation}
    \tilde{\phi}(z) = ze^{-i \chi} \prod_{i=1}^{w-1}\frac{z-z_i}{z^*_i(z-1/z^*_i)} ,
    \label{factor}
\end{equation}
where
$\{z_i\}$ are the zeros of ${\cal N}(z)$ or $\tilde{\phi}(z)$.
These zeros are located inside the unit circle in the complex plane.
As mentioned, they are also the conjugate of the eigenvalues 
of the survival operator ${\cal S}=\left( 1-\hat{D}\right) \hat{U}(\tau)$. 
This can be proven by applying the matrix determinant lemma 
to the characteristic polynomial of ${\cal S}$ \cite{Thiel2020D}, namely,
\begin{equation}
\begin{aligned}
    0 = 
    \text{det} \left[ \zeta \mathbb{1} - {\cal S} \right]
    = \text{det} [ \zeta \mathbb{1} - \hat{U}(\tau) 
    + \ket{0}\bra{0} \hat{U}(\tau) ] 
    = \text{det} [ \zeta \mathbb{1} - \hat{U}(\tau) ]
      \expval{[ \zeta \mathbb{1} - \hat{U}(\tau) ]^{-1}}{0}.
\end{aligned}
\end{equation}
The term $\expval{[ \zeta \mathbb{1} - \hat{U}(\tau) ]^{-1}}{0}$ 
can be spectrally decomposed as 
$\sum_{k=1}^w \sum_{l=1}^{g_k} |\bra{0}\ket{E_{kl}}|^2 \left[ 1/(\zeta -e^{-iE_k\tau}) \right]$,
which, equal to $0$, gives the eigenvalues of ${\cal S}$, $\{\zeta_i\}$, inside the unit disk, 
that are conjugate of the zeros of $\tilde{\phi}(z)$ 
(excluding the trivial zero $z=0$).
Namely, 
\begin{equation}
    \zeta_i = z_i^*.
\end{equation}


We note here that the mean hitting time $\expval{n}$ 
(for infinite measurements, i.e. $T_R = \infty$) can be computed by 
\begin{equation}\label{neqw}
    \expval{n} = {1\over 2\pi i} \oint_{|z|=1} 
    \partial_z \ln \left[ \tilde{\phi}(z) \right]\, {\rm d}z,
\end{equation}
which directly gives $\expval{n}=w$ using equation (\ref{factor}).
Namely, the mean $\expval{n}$ is identical to the number of zeros of $\tilde{\phi}(z)$, 
{\em inside the unit disk}. 


Substituting equation~(\ref{factor}) into equation~(\ref{inversion}) and using the residue theorem yield 
\begin{equation}\label{phi45}
    \phi_n = e^{-i \chi} \sum^{w-1}_{j=1} \left( z_j^{*} \right)^{n-1}
            \left( \frac{1}{z^*_j}-z_j \right) 
            \prod_{k\neq j} \frac{z_j^*(1/z^*_j-z_k)}{z^*_k-z^*_j}
            .
\end{equation}
%
Let $z_j=\rho_j \exp\left( i \theta_j \right)$, 
i.e. $\rho_j = |z_j| = |\zeta_i|$,
$\theta_j = \text{arg}(z_j)\in [0, 2\pi)$, 
and further simplification gives 
\begin{equation}
    F_n = \left| \phi_n \right|^2 
    = \sum_{j,k=1}^{w-1} 
    \frac{\alpha_j \alpha^*_k}{\beta_j \beta^*_k} \,
    \left( \rho_j \rho_k \right)^{n} e^{in \Theta_{jk}},
    \label{Fnab}
\end{equation}
where
$\Theta_{jk} = \theta_k-\theta_j \in [0, 2\pi)$, and 
\begin{equation}\label{para}
    \frac{\alpha_j}{\beta_j} 
    = \frac{\prod_{i} \left( 1/z^*_i \right)
    \left( 1/z^*_j-z_i \right)}{\prod_{i\neq j} \left( 1/z^*_j-1/z^*_i \right)}
    .
\end{equation}
Hence equation~(\ref{Fnab}) has $(w-1)^2$ terms. 
Due to the invariance under the switching between $j$ and $k$ in equation (\ref{Fnab}),
the fact that $F_n$ is real is guaranteed by the appearance of paired conjugate terms.
\begin{equation}\label{eq43}
    F_n \sim a_{\text{max}} \rho_{\text{max}}^{2n} = a_{\text{max}} \left| z_{\text{max}} \right|^{2n}
    = a_{\text{max}} \left| \zeta_{\text{max}} \right|^{2n},
\end{equation}
where $a_{\text{max}} = \left| {\alpha_{\text{max}} / \beta_{\text{max}}} \right|^2$.
Using equation (\ref{para}), and $\rho_{\text{max}}=|\zeta_{\text{max}}| = 1-\varepsilon \to 1$,
we have
\begin{equation}\label{eq44}
    a_{\text{max}} = \left| {\alpha_{\text{max}} \over \beta_{\text{max}}} \right|^2 
    = \left| {1 \over z_{\text{max}}^*} \left( {1 \over z_{\text{max}}^*} - z_{\text{max}} \right) \right|^2
    \sideset{}{'}\prod_i 
    \left| { 1/z^*_{\text{max}}-z_i \over \left( z^*_i \right) \left( 1/z^*_{\text{max}}-1/z^*_i \right)} \right|^2
    \sim 
    \left( 1 - |z_{\text{max}}|^{-2} \right)^2 
    \sideset{}{'}\prod_i
    \left| { 1-z_i \over z_i^* -1} \right|^2
    \sim \left( 1 - \rho_{\text{max}}^{2} \right)^2
    ,
\end{equation}
where $\prod^\prime_i$ means multiplication over all ii except for $z_i=z_\text{max}$.
Therefore, we get a universal formula for $F_n$'s tail, 
in the vicinity of the transition or phase factors matching, namely,
\begin{equation}\label{eq53}
    F_n \sim \left( 1-\rho_{\text{max}}^{2} \right)^2 \rho_{\text{max}}^{2n} 
    = \left( 1 - |\zeta_{\text{max}}|^{2} \right)^2 |\zeta_{\text{max}}|^{2n},
\end{equation}
which confirms rigorously the validity of equation (8).
We have assumed that a gap exists between the maximum $|\zeta_{\text{max}}|$ and other zeros of ${\cal N}(z)$ in the system. 
Note: All along we assumed that the Hilbert space is finite, otherwise the spectrum becomes degenerate. 
Finally, with equation (\ref{eq53}) we derive our main results in equations (6,7).
We want to note again that $\zeta_{\text{max}}$ is unique in our work.

\section*{3. Dependence of restart uncertainty relation on system size}
We now discuss the relation between the restart uncertainty principle 
and the size of the system. 
Recall that we use the notation $H\ket{E_{k,l}} = E_k \ket{E_{k,l}}$ 
where $H$ is the Hamiltonian, $l$ is an index that accounts for possible degeneracy of the energy level. 
Then when two energy phases match $\exp (- i E_{-} \tau) \sim \exp ( - i E_+ \tau)$ 
for a pair of energies $E_+$ and $E_{-}$, 
where $\tau$ is the sampling time, 
we find a resonance in the mean number of measurement till the first detection. 
In particular, using the equations (13) and (14) in the main text, 
we state the uncertainty related to system energy,
\begin{equation}\label{eq13}
    \expval{n}_\text{Con} 
    = w - 
    \left[ 1+ \lambda T_R (\widetilde{\Delta E \tau})^2 \right]
    \exp 
    \left[ -\lambda T_R (\widetilde{\Delta E \tau })^2 \right],
\end{equation}
\begin{equation}\label{eq14}
    \expval{n_R} 
    = w - 
    \exp 
    \left[ -\lambda T_R (\widetilde{\Delta E \tau })^2 \right],
\end{equation}
where $\expval{n}_\text{Con}$ is the conditional mean,
$\expval{n_R}$ is the restarted mean and $w$ is the topological number 
which is determined by the distinct energy eigenvalues of the system.
Later we will only focus on the restarted mean 
since similar behaviors are found for the conditional mean.
The parameters 
$\lambda =  {p_+p_- / (p_+ + p_-)^3}$ 
with the overlaps $p_\pm = \sum_l^{g_\pm} \abs{\braket{0}{E_{\pm,l}}}^2$ 
($g_\pm$ is the degeneracy of the energy level $E_\pm$, and the location of the target $x_{\rm d} = 0$ as in the main text), 
and 
\begin{equation}
\widetilde{\Delta E \tau } := \tau | E_+ - E_- | \mod 2 \pi. 
\end{equation}
Hence, 
we need to find out how the size of a system will affect its energy levels 
and energy eigenstates (which determine the overlaps $p_\pm$ and $\lambda$).
Since energies depend on system size, 
so will the resonances, however, additionally $\lambda$ is also generally size-dependent.
This implies rich types of physical behaviors as system size is changed.

Without delving into details, we have summarized in Table \ref{uncTab}, 
the values of parameters in equations (\ref{eq13}) and (\ref{eq14}), for different graphs, 
with the resonance chosen at 
$\exp (- i E_{\rm \max} \tau) \sim\exp ( - i E_{\rm min} \tau)$, 
{where $E_{\rm \max}$ and $E_{\rm \min}$ are the maximum and minimum of energies 
of the system respectively}.
It is clearly shown that different graph structures lead to various relations
between the width of transitions and the system size $L$.
%
\begin{figure}[h]
\begin{center}
\includegraphics[width=0.7\linewidth]{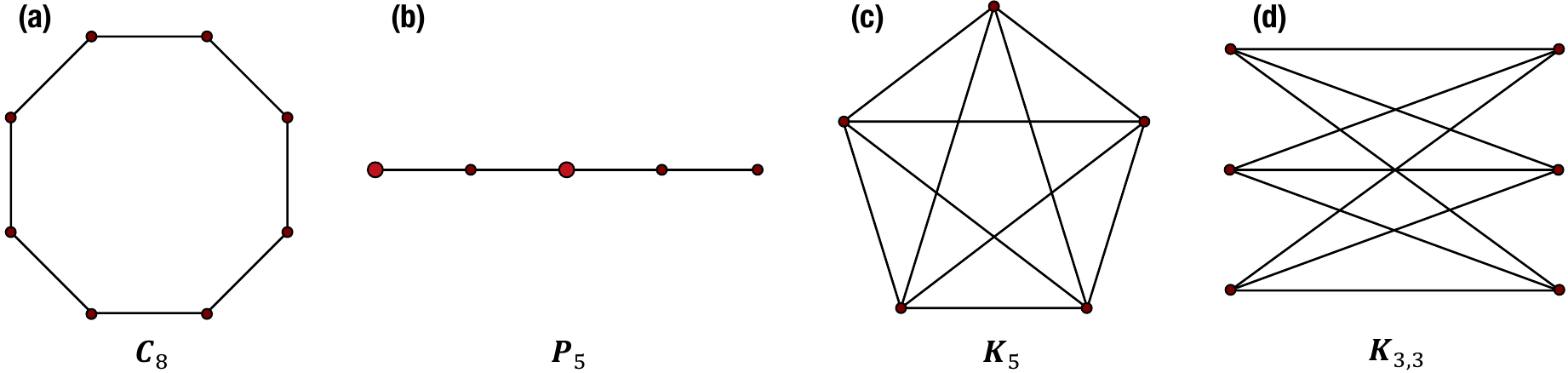}
\end{center}
\caption{
Schematics for the graphs under investigation.
From the left to the right are the examples:
(a) ring of size $L$, $C_L$, in the figure $L=8$,
(b) finite segment of size $L$, $P_L$, in the figure $L=5$,
(c) complete graph of size $L$, $K_L$, in the figure $L=5$,
(d) complete bipartite graph $K_{m,n}$, in the figure $m=n=3$, and the size is $L = m+n = 6$.
We will verify our theory using different sizes.
For graph (b), the target site will be chosen at the end or the middle
(marked with larger vertices),
which leads to non-identical resonance widths. 
}
\label{fig:examgraphs}
\end{figure}
\begin{table*}[t]
\centering
\caption{
The winding number $w$, 
the maximal difference between energies 
$\Delta E_{\rm m} = E_{\rm max} - E_{\rm min}$,
used in $\widetilde{\Delta E \tau} = (\tau \Delta E_{\rm m} \mod 2 \pi)$,
and the parameter $\lambda = p_+ p_- / (p_+ + p_-)^3$,
for different graphs with $L$ vertices, 
including even rings $C_L$, complete graphs $K_L$,
finite segments $P_L$, and complete bipartite graphs $K_{L/2,L/2}$.
Only for the segment, i.e. the $P_L$ graph, 
the location of the target, denoted $x_{\rm d}$, 
is important.
}
\label{uncTab}
\sisetup{detect-weight,mode=text}
\renewrobustcmd{\bfseries}{\fontseries{b}\selectfont}
\renewrobustcmd{\boldmath}{}
\newrobustcmd{\B}{\bfseries}
%
\begin{tabular}{ cccccccccccccccccccc } 
 \Xhline{2\arrayrulewidth}
 \multicolumn{4}{c}{ \B \,\,\,Graph\,\,\,} & 
 \multicolumn{4}{c}{\makecell{ ${\bm C_L}$, $L$ is even}} & 
 \multicolumn{4}{c}{\makecell{ ${\bm K_L}$}} & 
 \multicolumn{4}{c}{\makecell{ ${\bm P_L}$, $L$ is odd}} & 
 \multicolumn{4}{c}{\makecell{ ${\bm K_{{L\over2},{L\over2}}}$}}  \\~\\
 \multicolumn{4}{c}{${\bm w}$} & 
 \multicolumn{4}{c}{\,\,${L / 2 +1}$\,\,}  & 
 \multicolumn{4}{c}{\,\,$2$\,\,}  & 
 \multicolumn{4}{c}{\makecell{$L$, $x_{\rm d}=1$; \\
 ${(L+1) / 2}$, $x_{\rm d} = {(L+1) / 2}$}}  & 
 \multicolumn{4}{c}{\,\,$3$\,\,} \\~\\
 \multicolumn{4}{c}{${\bm \Delta E_{\rm m}}$} & 
 \multicolumn{4}{c}{$4\gamma$} & 
 \multicolumn{4}{c}{$1$} & 
 \multicolumn{4}{c}{$4\gamma \cos\left[ {\pi / (L+1)} \right]$} & 
 \multicolumn{4}{c}{$\gamma L$} \\~\\
 %
 \multicolumn{4}{c}{${\bm \lambda}$} & 
 \multicolumn{4}{c}{${L / 8}$} & 
 \multicolumn{4}{c}{${(L-1) / L^2}$} & 
 \multicolumn{4}{c}{\makecell{${(L+1)^3 / 16\pi^2}$, $x_{\rm d}=1$; \\
 ${(L+1) / 16}$, $x_{\rm d}={(L+1) / 2}$}} & 
 \multicolumn{4}{c}{${L / 8}$} \\
 %
 \Xhline{2\arrayrulewidth}
\end{tabular}
\end{table*}
%
%

\begin{figure}[h]
\begin{center}
\includegraphics[width=0.8\linewidth]{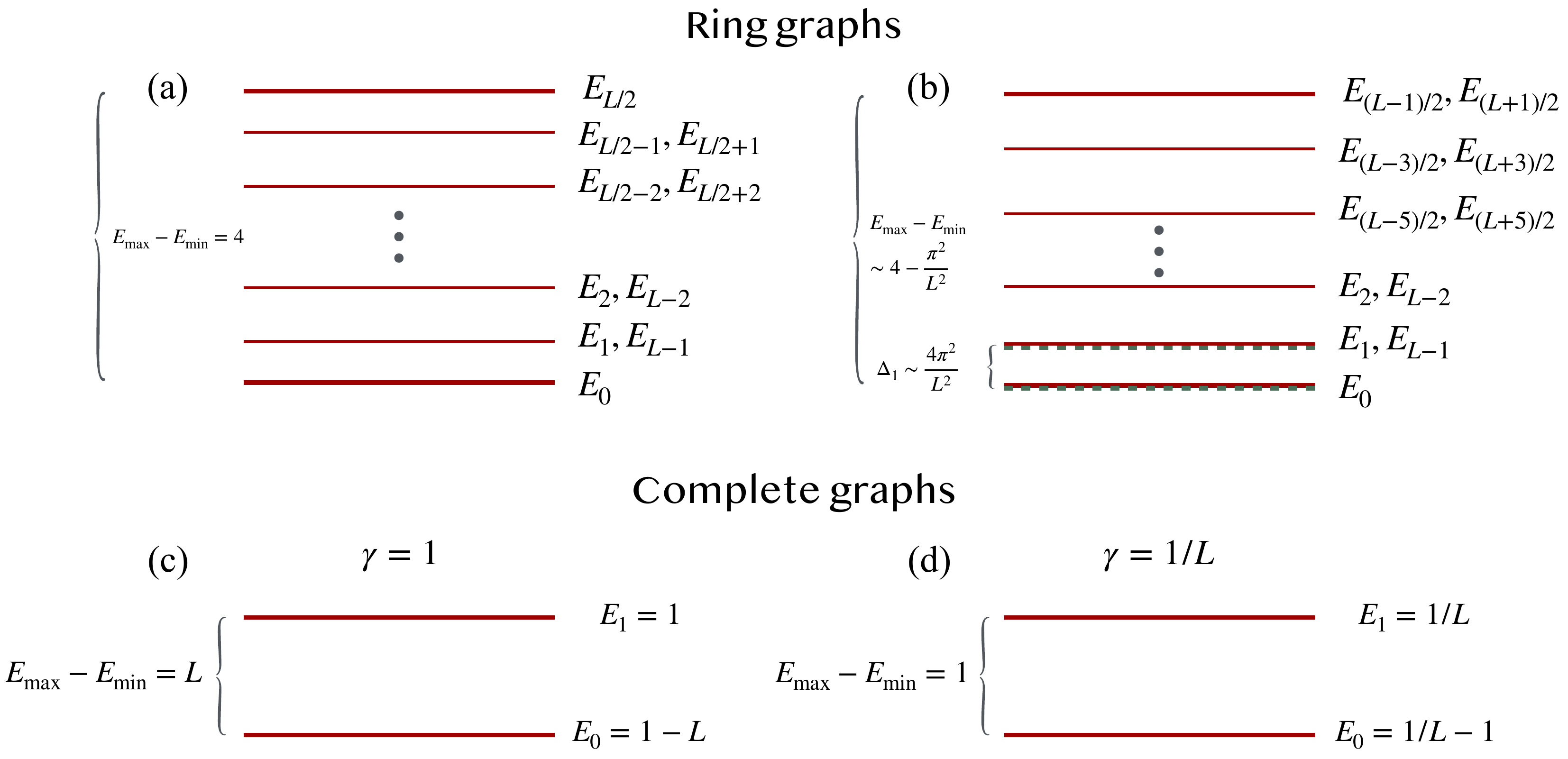}
\end{center}
\caption{
The energy levels of ring graphs and complete graphs. 
In (a) we present the case of even $L$, while in (b) $L$ is odd.
We consider the resonance related to the largest energy
and the lowest energy (ground state energy),
which we called the min-max condition. 
As a second option
we choose the ground state energy and the first excited state energy.
The dispersion relation for rings is $E_k = -2 \gamma \cos(2\pi k /L)$ 
with $k=0,1,2,\dots,L-1$ and $\gamma=1$.
Here $\gamma$ is the hopping amplitude between nodes, 
namely $H$ is the adjacency matrix of the graph multiplied by $\gamma$.
For complete graphs (subplot (c)), the energies are $1$ and $1-L$.
As typically used in literature, 
the hopping rate $\gamma$ is set as inversely proportional 
to the number of edges of each vertex, 
see subplot (d) where the energy difference is $E_{\rm max}- E_{\rm min} = 1$.
}
\label{fig:enlring}
\end{figure}

{\bf Ring models.} We start with the ring model (Figure \ref{fig:examgraphs}(a)),
which is used for demonstration purposes in the manuscript. 
Energies of the ring model of size $L$ are 
$E_k = - 2 \gamma \cos \theta_k$ 
with $\theta_k = {2 \pi k / L}$ and $k=0,1,2,\dots, L-1$ (see equation (17) in the text), 
and overlaps are $\left| \braket{x}{E_k}\right|^2 = 1/L$ for any node $x$ , 
the broadening can be easily associated with the system size $L$ (assuming even $L$).
See Figure \ref{fig:enlring}(a-b) for a schematics of its energy structures,
where the parity of $L$ plays a role.
We start the discussion where the pair of energies is $E_{\rm max}$ and $E_{\rm min}$, 
and then consider the case when we chose the energy difference 
between the the ground state and the first excited state 
(this is based on odd ring, otherwise the transition will be $w \to w - 2$ 
which is left for future study). 

\begin{figure}[t]
\begin{center}
\includegraphics[width=0.45\linewidth]{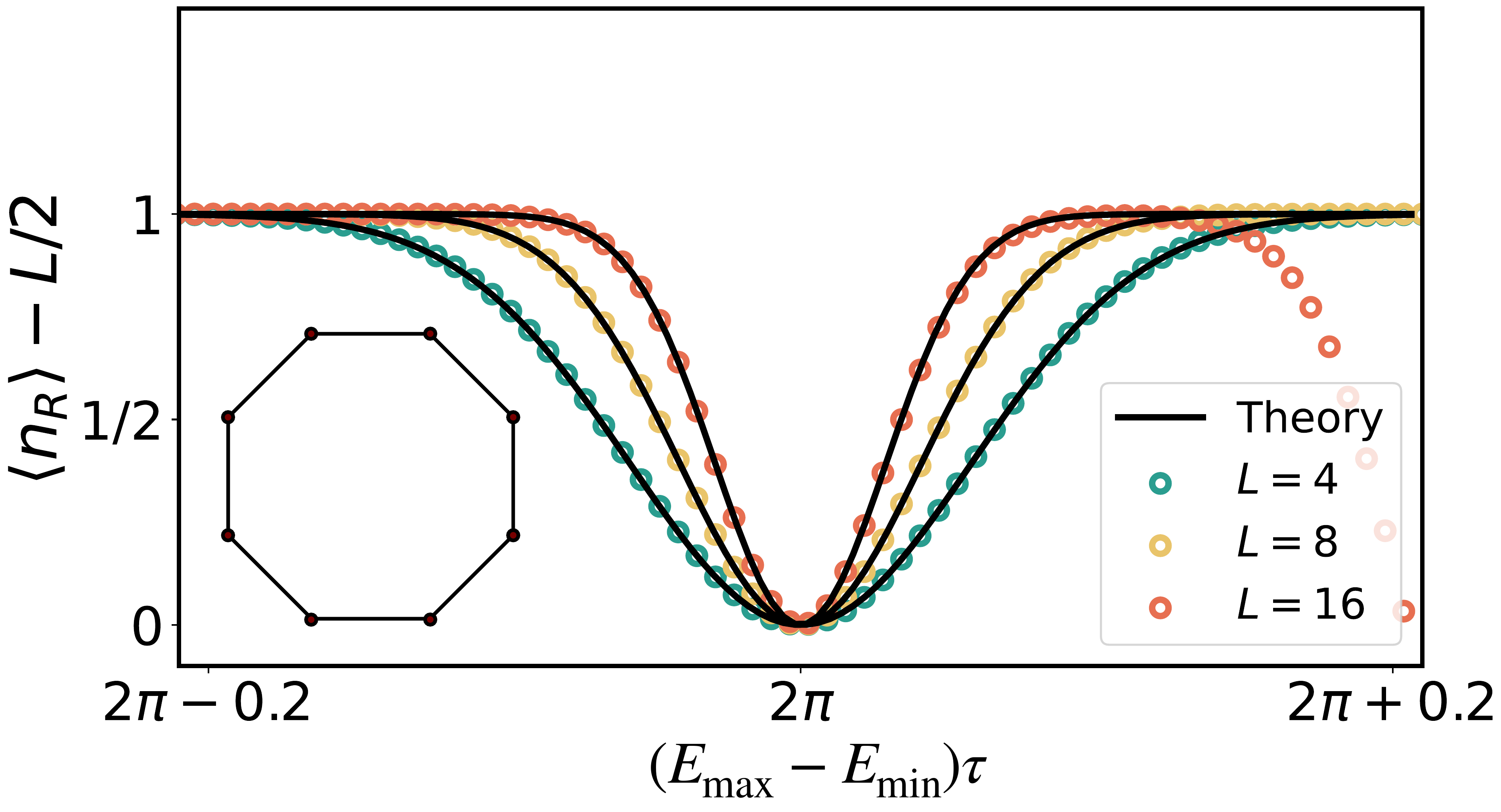}
\end{center}
\caption{ Restarted mean hitting times 
versus $(E_{\rm max} - E_{\rm min}) \tau$, 
for ring graphs of various sizes $L$ (see inset for an example).
The resonances become narrower as we increase the size of the system. 
Recall, that difference between the largest and ground-state energies, 
$E_{\rm max} - E_{\rm min} = 4\gamma$, is size-independent, and we choose $\gamma=1$. 
We shift the mean by $L/2$, to focus on the width of the transition. 
The numerical results are obtained with equations (2-4) in the main text, 
and this perfectly matches our theory, 
see equation (\ref{eq4}).
The deviation on the right for $L=16$ is caused by the proximity of another resonance.
$T_R=300$ is used here. 
Similar results for $\expval{n}_{\rm Con}$ were also tested, 
and not presented hereinafter.
}
\label{fig:ring0}
\end{figure}

For the resonance between the ground state and the highest energy state, 
where phase factors $\{e^{-i2\gamma\tau}, e^{i2\gamma\tau} \}$ merge,
we have $\widetilde{\Delta E \tau } = \tau \Delta E_{\rm m} \mod 2\pi 
= 4\gamma\tau \mod 2\pi$
(and now we set $\gamma$ as $1$),
and $\Delta E_{\rm m} = E_{\rm max} - E_{\rm min}$.
Thus, for even $L$,

%
\begin{equation}
    \begin{split}
        \langle n_R\rangle 
        &= 
        w
        - 
        \exp[- L(\widetilde{\Delta E \tau })^2  T_R /8 ], 
    \end{split}
    \label{eq4}
\end{equation}
where $w=(2 + L)/2$.
We note that for odd ring, $w=(L+1)/2$, 
hence $\Delta E_{\rm m}$ is $4-\pi^2/L^2$.
See Figure \ref{fig:enlring}(b).
Thus, with $L$ increasing, the broadening of the transition will be narrower, 
for all the rings with odd or even number of nodes.
See Figure \ref{fig:ring0}, where we present numerical confirmation for even rings.
%
%
%

However, if we consider the resonance related to 
the ground state and the first excited state, for the odd rings, 
which leads to the transition $w\to w-1$,
the $L$ dependence of the energy difference will be distinct.
In this case it follows that
$\widetilde{\Delta E \tau } = \tau \Delta_1 \mod 2\pi 
= (E_{\rm 1st} - E_{\rm g})\tau \mod 2\pi$
with $E_{\rm 1st} - E_{\rm g} \sim 1/L^2$.
i.e. the energy difference shrinks when the system size $L$ grows 
(see Figure \ref{fig:enlring}(b)).
The parameter $\lambda$ is still proportional to $L$,
and then the term $\lambda (\widetilde{\Delta E \tau })^2$ 
is proportional to $1/L^3$, when $\tau$ is tuned close to the resonance.
Hence this will result in an increasing width of the resonance
as we increase the size $L$.

Now it is readily realized that
the system size $L$ has various ways of entering the expressions 
for energy levels and eigenstates. 
In the context of quantum walks on graphs, 
this means that the graphs, on which we dispatch quantum walkers, matter.
Different graph structures lead to different dispersion relations $E_k$, 
as well as the corresponding eigenvectors $\ket{E_k}$.
To explore how $L$ determines $\lambda$ and $\widetilde{\Delta E \tau}$, 
we checked other graphs.

\begin{figure}[t]
\begin{center}
\includegraphics[width=0.45\linewidth]{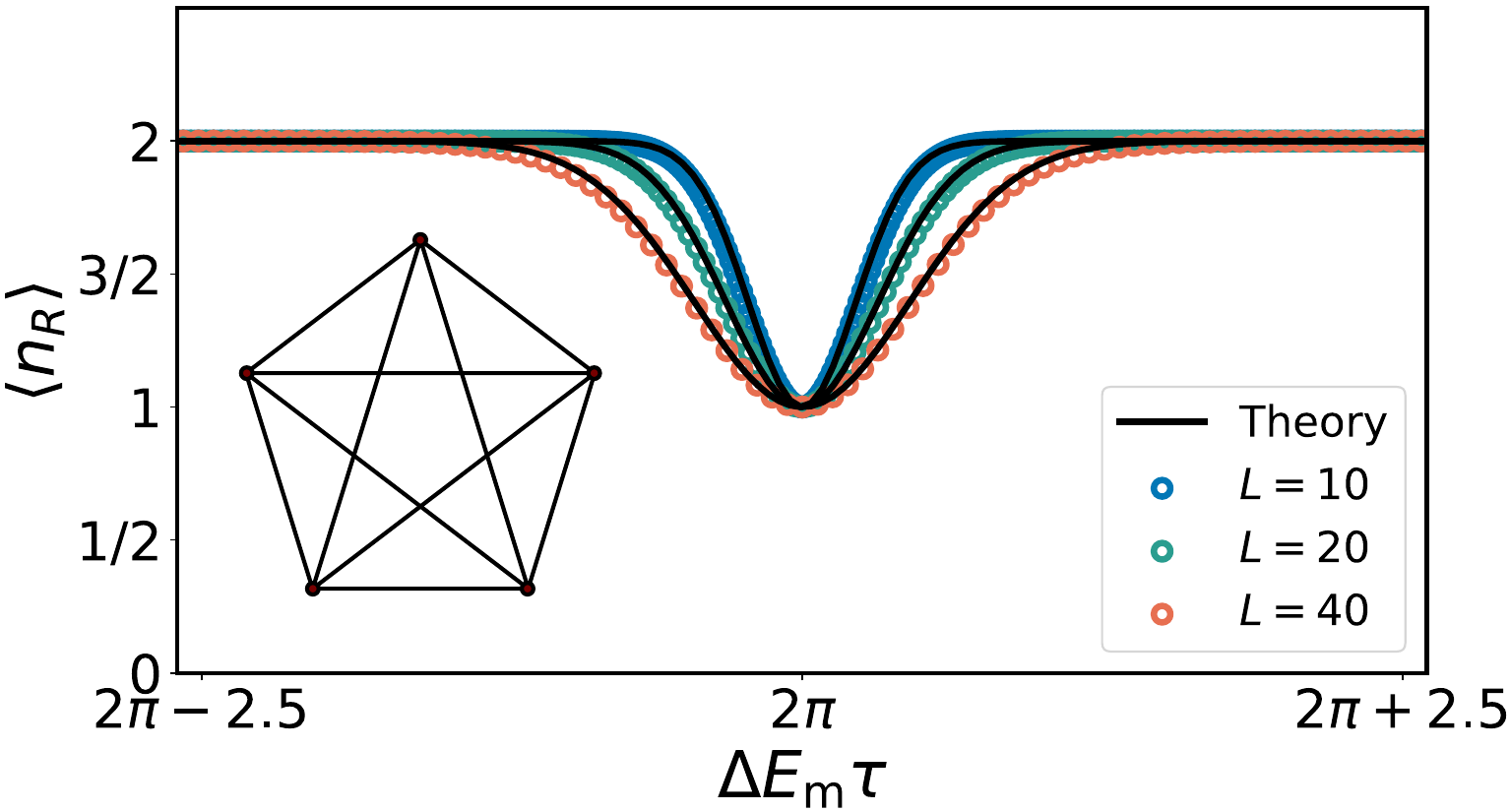}
\end{center}
\caption{
Restarted mean hitting time 
for complete graphs of different sizes 
(see inset for an example). 
The numerical results are obtained with equations (2-4) in the main text, 
and the theory is computed with equation (\ref{eq5}).
Here $\Delta E_{\rm m} = 1$ and we used $\gamma=1/L$ for a fair comparison. 
Unlike Figure \ref{fig:ring0}, the broadening becomes wider as the system size $L$ increases.
Here we used $T_R=100$.
}
\label{fig:cg1}
\end{figure}
\begin{figure}[t]
\begin{center}
\includegraphics[width=0.45\linewidth]{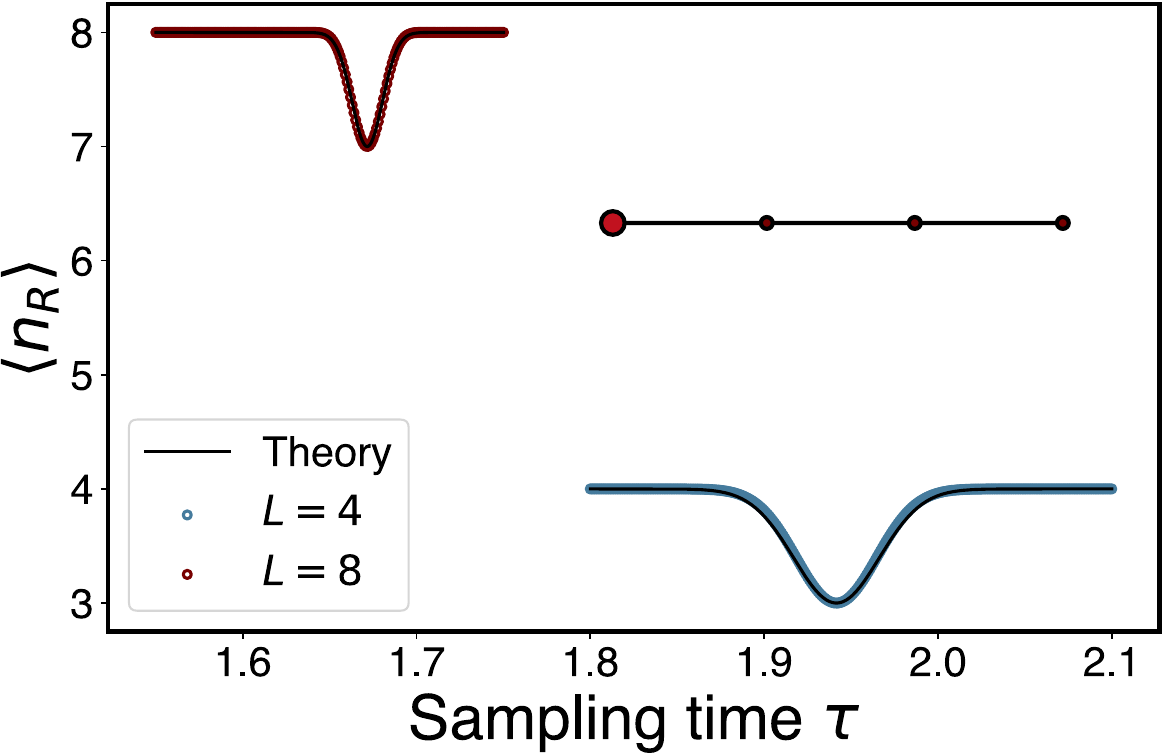}
\end{center}
\caption{
Restarted mean hitting time 
for finite segments of different sizes.
Repeated measurements are made on the leftmost node 
(see schematics in the inset, where the larger circle points to the measured node).
The numerical results are obtained with equations (2-4) in the main text, 
and the theory is computed with equation (\ref{line1}).
We see that the broadening becomes narrower as the system size $L$ increases.
Here $T_R=100$. 
}
\label{fig:lineend}
\end{figure}
\begin{figure}[t]
\begin{center}
\includegraphics[width=0.45\linewidth]{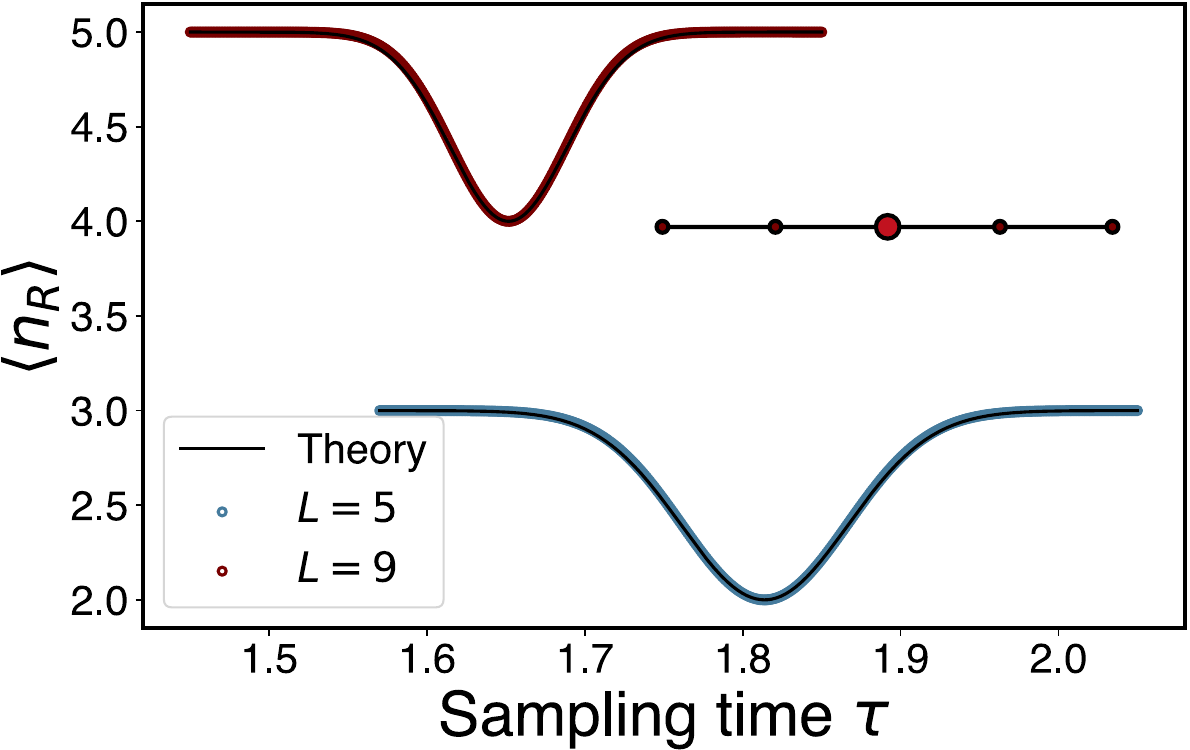}
\end{center}
\caption{
Restarted mean hitting time 
for finite segments of different sizes, 
here the target is set at $x_\text{d}=(L+1)/2$. 
The numerical results are obtained with equations (2-4) in the main text, 
and the theory is computed with equation (\ref{line2}).
We see that the broadening becomes narrower as the system size $L$ increases.
$T_R=40$ is used.
Here measurement is performed on the middle node, see inset.
}
\label{fig:linemid}
\end{figure}
%

{\bf Complete graph models.} One example is the complete graph, in which each vertex is connected to every other vertex. 
See Figure \ref{fig:examgraphs}(c). 
Specifically, the governing Hamiltonian in matrix form, 
has all elements equal $-\gamma$ except for the diagonal. 
To achieve a fair comparison, the hopping rate is usually chosen as $\gamma=\gamma_0/L$, 
and we set $\gamma_0=1$ here.
There are merely two energy levels, $E_0=\gamma(1-L)$ and $E_1=\gamma$, 
and the eigenstate corresponding to $\gamma(1-L)$, 
or the ground state is $(1,1,1,\dots,1)/\sqrt{L}$, 
hence the overlaps, for any initial/target state, are $p_+ = 1/L$ and $p_- = (L-1)/L$. 
This further leads to $\lambda = p_+ p_- /(p_+ + p_-)^3 =(L-1)/L^2 \sim 1/L$ 
as $L$ is large.
Hence with equations (13) and (14) in the text, we have for large $L$,
\begin{equation}
    \begin{split}
        \langle n_R \rangle &\sim 
        w
        - 
        \exp[- (\widetilde{\Delta E \tau })^2 T_R/L ], 
    \end{split}
    \label{eq5}
\end{equation}
where $w=2$ and $\widetilde{\Delta E \tau} 
= \tau \Delta E_{\rm m} \mod  2 \pi = L\gamma \tau \mod  2 \pi$.
From here we see that if choosing $\gamma$ independent on $L$, say $\gamma=1$,
we will have $\widetilde{\Delta E \tau} = L \tau \mod  2 \pi$ 
since the energy difference becomes $L$,
as shown in Figure \ref{fig:enlring}(c).
Then equation (\ref{eq5}) becomes 
$\langle n_R \rangle \sim w - \exp[- L(\tau - 2\pi/L)^2 T_R ]$ when $\tau\simeq 2\pi/L$,
indicating again a decreasing width of resonance as $L$ grows.

As mentioned above, we could also choose $\gamma=1/L$ as done in the literature of  quantum walks, 
which leads the energy difference to $\Delta E_{\rm m} =1$, 
as shown in Figure \ref{fig:enlring}(d).
Then we have $\langle n_R \rangle \sim w - \exp[- (\tau \mod 2\pi)^2 T_R/L]$,
suggesting an increasing width of resonance with the system size $L$ increasing.
%
See Figure \ref{fig:cg1} for the graphic demonstration. 

{\bf  Linear segments, bipartite graphs.}
We also checked linear segments,
and complete bipartite graphs $K_{L/2,L/2}$ 
(See Figure \ref{fig:examgraphs}(c) and (d)), 
for both of them and around the resonance where 
$\{e^{-i\tau E_\text{max}}, e^{-i\tau E_\text{min}} \}$ merge, 
the prefactor of $(\tau-\tau_c)^2$ (with $\tau_c$ the resonance $\tau$) 
is proportional to $L^3T_R$, suggesting again narrower broadening as $L$ grows.
More concretely, for a line of size $L$, the Hamiltonian is 
$H = -\gamma \sum_{x=1}^{L-1} (\ket{x} \bra{x+1} + \ket{x+1} \bra{x})$,
whose energy levels are $E_k = -2\gamma \cos[k\pi/(L+1)]$ with $k=1,2,\dots, L$,
and the corresponding eigenvectors are 
$\ket{E_k} = \sqrt{2/(L+1)} \sum_{j=1}^L \sin[k\pi j/(L+1)] \ket{j}$.
Hence there are $L$ distinct energies (no degeneracy), 
with the largest (lowest) energy $E_L=2\gamma \cos[\pi/(L+1)]$ ($E_1=-E_L$),
and the overlaps, for certain target site $x_\text{d}$, are
$p_k = |\bra{x_\text{d}}\ket{E_k}|^2 = [2/(L+1)] \sin^2[k\pi x_\text{d}/(L+1)]$.
Without loss of generality, assuming odd $L$,
we consider $x_\text{d}$ at either end of the line, or the middle of the line,
namely, $x_\text{d}=1$ or $x_\text{d}=(L+1)/2$, leading to 
$p_k = |\bra{1}\ket{E_k}|^2 = [2/(L+1)] \sin^2[k\pi /(L+1)]$,
or
$p_k = |\bra{{L+1 \over 2}}\ket{E_k}|^2 = [2/(L+1)] \sin^2(k\pi /2) = [1-(-1)^k] / (L+1)$,
respectively.
We find that the $p_k$'s are non-zero for the former case, 
while for the latter, $x_\text{d}=(L+1)/2$, there appear $p_l=0$ when $l$ is even.
This leads to different winding numbers for the two cases, 
since $w$ is equal to the number of distinct phases $e^{-iE_k\tau}$ 
associated with non-zero $p_k$.
Hence $w=L$ for the case $x_\text{d}=1$,
and $w=(L+1)/2$ for the case $x_\text{d}=(L+1)/2$.
We now focus on the resonance 
where phases $\{ e^{-i\tau E_1}, e^{-i\tau E_L} \}$ merge at
$\tau_c=2\pi/|E_1-E_L| = \pi/2\gamma \cos[\pi/(L+1)]$,
which is the smallest resonance $\tau$ except for $\tau=0$.
For the target at one end of the line, $x_\text{d}=1$,
the corresponding overlaps to $E_1$ and $E_L$ are 
$p_1 = p_L = [2/(L+1)] \sin^2[\pi /(L+1)] \approx 2\pi^2/(L+1)^3$, 
with the approximation valid when $L$ is large.
For the case $x_\text{d}=(L+1)/2$, we have 
$p_1 = p_L = 2/(L+1)$.
Therefore, for $x_\text{d} =1$, namely the end node on the segment,
equations (13) and (14) for large $L$ become
\begin{equation}\label{line1}
    \begin{split}
        \langle n_R \rangle &= 
        w - 
        \exp \left\{
        - {(L+1)^3 \over 16\pi^2} 
        \left[ 4 \cos({\pi \over L+1}) \tau \mod 2 \pi \right]^2 
        T_R 
        \right\} , 
    \end{split}
\end{equation}
where $w = L$. 
And for the target site at the middle of the line, $x_\text{d} = (L+1)/2$, we have
\begin{equation}\label{line2}
    \begin{split}
        \langle n_R \rangle &= 
        w - 
        \exp \left\{
        - {L+1 \over 16} 
        \left[ 4 \cos({\pi \over L+1}) \tau \mod 2 \pi \right]^2 
        T_R 
        \right\}, 
    \end{split}
\end{equation}
where $w=(L+1)/2$.
Clearly, 
these expressions exhibit a different dependence on system size.
See Figures \ref{fig:lineend} and \ref{fig:linemid} for numerical confirmation,
where the theory works well 
and predicts the narrowing of broadening of resonances as the system becomes larger. 
%

\begin{figure}[t]
\begin{center}
\includegraphics[width=0.45\linewidth]{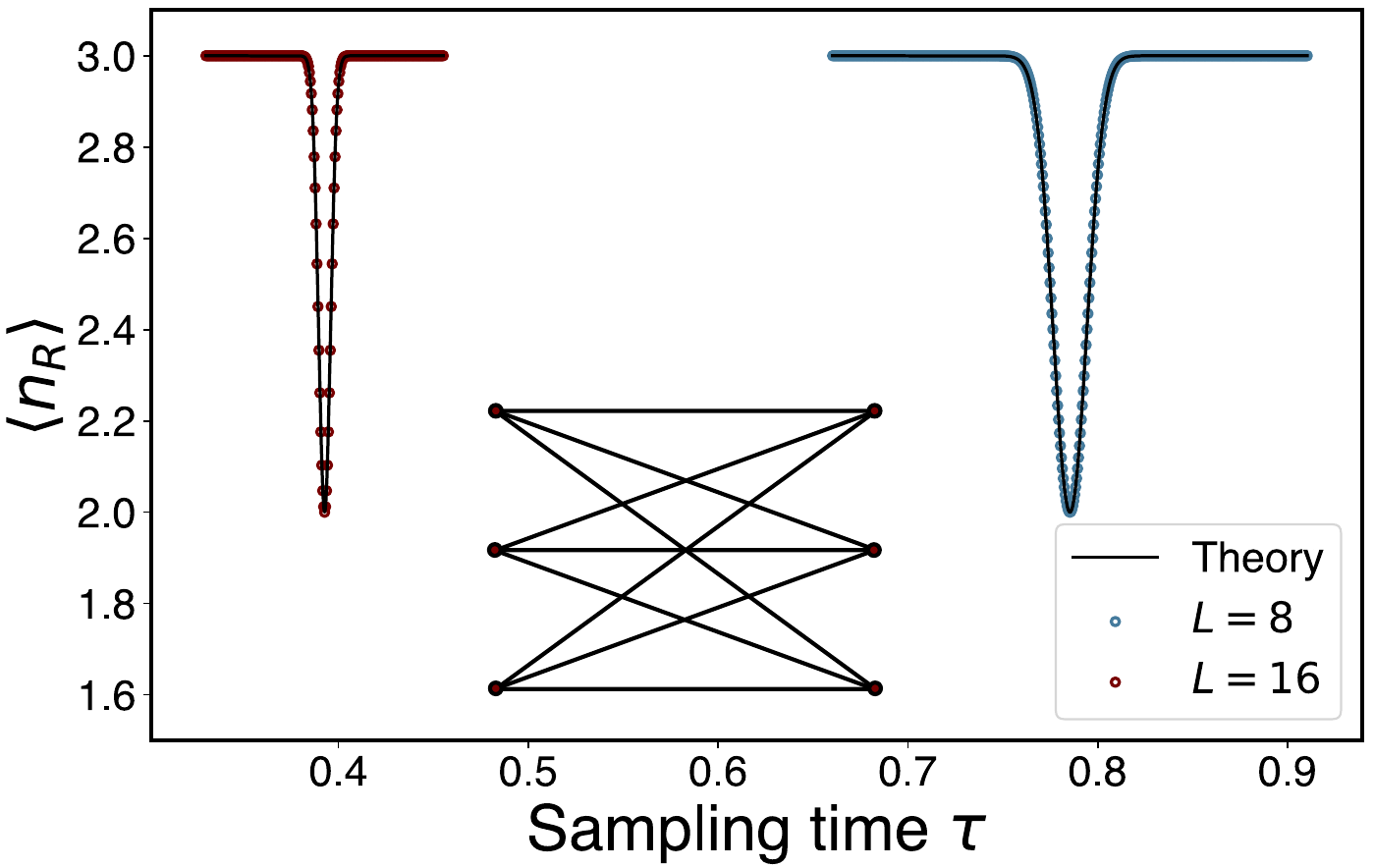}
\end{center}
\caption{
Restarted mean hitting time for complete bipartite graphs $K_{L/2,L/2}$ 
with different $L$ (see an example in Figure \ref{fig:examgraphs}d).
The numerical results are obtained with equations (2-4) in the main text, 
and the theory is computed with equation (\ref{cb}).
We see that the broadening becomes narrower as the system size $L$ increases.
Here $T_R=100$.
}
\label{fig:combip}
\end{figure}

Another example is a complete bipartite graph, also called a complete bi-colored graph, 
usually denoted by $K_{l,m}$, see Figure \ref{fig:examgraphs}(d). 
The vertices of the graph can be decomposed into two disjoint sets, 
containing $l$ and $m$ elements, respectively,
such that {\em no} two vertices within the {\em same} set are connected by an edge,
and every pair of vertices from {\em different} sets are connected.
See Figure \ref{fig:examgraphs}(d) for schematics of $K_{3,3}$.
Here we use $K_{L/2,L/2}$ to demonstrate the influence of size $L$ 
on the restart uncertainty relation.
The Hamiltonian governing a quantum walk on such a graph is
$H = -\gamma 
\begin{bmatrix}
	O &C \\
	C &O
\end{bmatrix}
$
with $C$ a ${L\over 2} \times {L\over 2}$ matrix with all elements as $1$.
The energy levels are 
$\gamma \{ -L/2, 0, L/2 \}$.
The eigenvectors corresponding to the lowest and largest energies are
$\ket{E_0} = \left(-1, -1, -1, \cdots, -1,-1,-1, \cdots \right)^T/\sqrt{L}$,
and $\ket{E_2} = \left(-1, -1, -1, \cdots, 1,1,1, \cdots \right)^T/\sqrt{L}$.
This gives, around the resonance where $\{ e^{-i\tau E_0}, e^{-i\tau E_2} \}$ merge,
the overlaps $p_0 = p_2 = 1/L$, for any node as the target site.
Hence the parameters are straightforwardly calculated, namely
$\lambda = L/8$, and $\widetilde{\Delta E \tau} = \tau L \mod 2\pi$ 
($\gamma$ is set as $1$).
Thus the statistical measures of mean hitting time around the resonance is
%
%
\begin{equation}\label{cb}
    \begin{split}
        \langle n_R \rangle &= 
        w - 
        \exp[
        - L 
        (\widetilde{\Delta E \tau})^2 
        T_R /8
        ], 
    \end{split}
\end{equation}
where $w = 3$.
In the vicinity of the resonance $\widetilde{\Delta E \tau}\simeq 0$, 
$(\tau L \mod 2\pi)^2$ becomes $(\tau L - 2\pi)^2$,
thus, we get
\begin{equation}\label{cb}
    \begin{split}
        \langle n_R \rangle &= 
        w - 
        \exp[
        - L^3 
        (\tau - 2\pi/L)^2 
        T_R /8
        ], 
    \end{split}
\end{equation}
In Figure \ref{fig:combip} we present the numerical results 
calculated with equations (2-4) in the main text, 
and our theory agrees excellently with the numerics.
Therefore, as theoretically predicted and numerically seen,
the increasing system size leads to more abrupt transitions of the mean hitting times,
namely the resonance is narrowed as we increase $L$.

\section*{4. Effects of non-precise sampling time and restart time}\label{append:B}
We elaborate here how we implement the randomness of $\tau$ and $T_R$ in our problem,
and what we witness for their effects upon the 
restart uncertainty relation and the broadening of resonance phenomenon.

%
\begin{figure}[t]
\begin{center}
\includegraphics[width=0.45\linewidth]{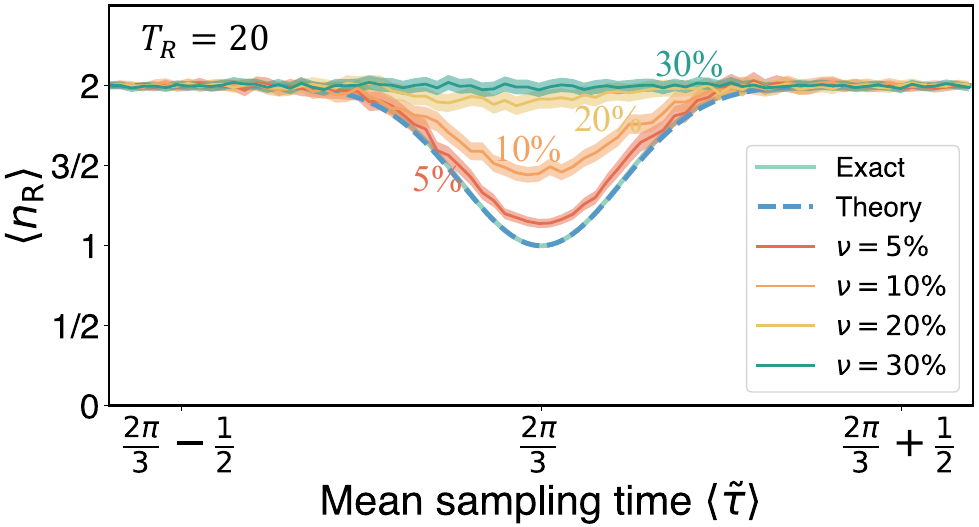}
\end{center}
\caption{ Mean hitting time versus the mean sampling time $\langle \tilde{\tau} \rangle $, for the three-site ring model 
with varying inaccuracy levels in the evolution time \(\tau\) 
and fixed \(T_R = 20\). 
Utilizing the Monte Carlo method with \(30,000\) realizations (conducted with {\em Python}), 
we find that as the fluctuations of $\tau$ increase, 
the resonances are progressively diminished, 
yet the topological number $\langle n_R \rangle = 2$, far from the resonance, remains unaffected and exhibits robustness.
}
\label{fig3}
\end{figure}
\begin{figure}[t]
\begin{center}
\includegraphics[width=0.45\linewidth]{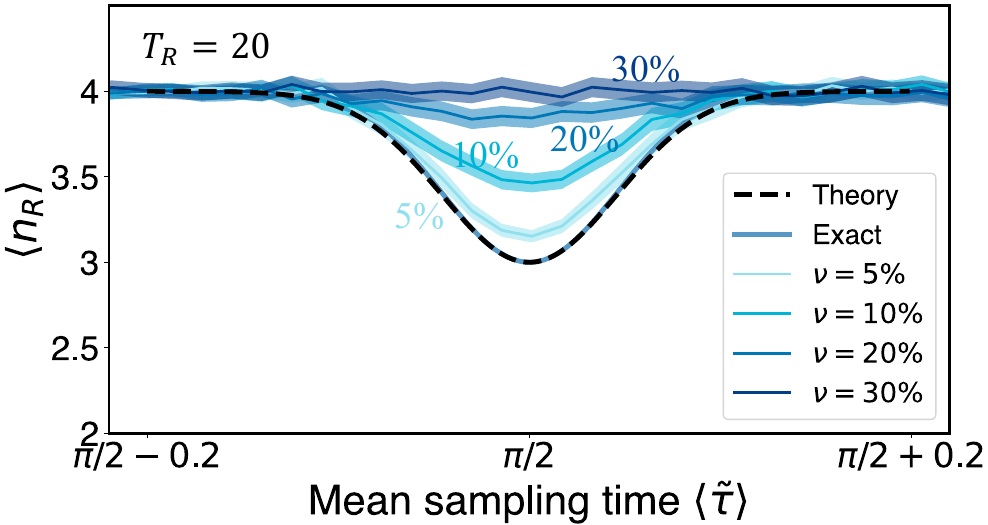}
\end{center}
\caption{ Mean hitting time versus the mean sampling time $\langle \tilde{\tau} \rangle $, for the benzene-type ring model, 
with varying inaccuracy levels in the evolution time \(\tau\) 
and a fixed restart time \(T_R = 20\).
Utilizing the Monte Carlo method, 
we simulated the first detection process with restarts
across \(30,000\) realizations. 
Our results demonstrate that as the fluctuations of $\tau$ increase, 
the resonances are progressively diminished, 
yet the topological number $\langle n_R \rangle = 4$ remains unaffected and exhibits robustness.
}
\label{fig4}
\end{figure}
\begin{figure}[t]
\begin{center}
\includegraphics[width=0.45\linewidth]{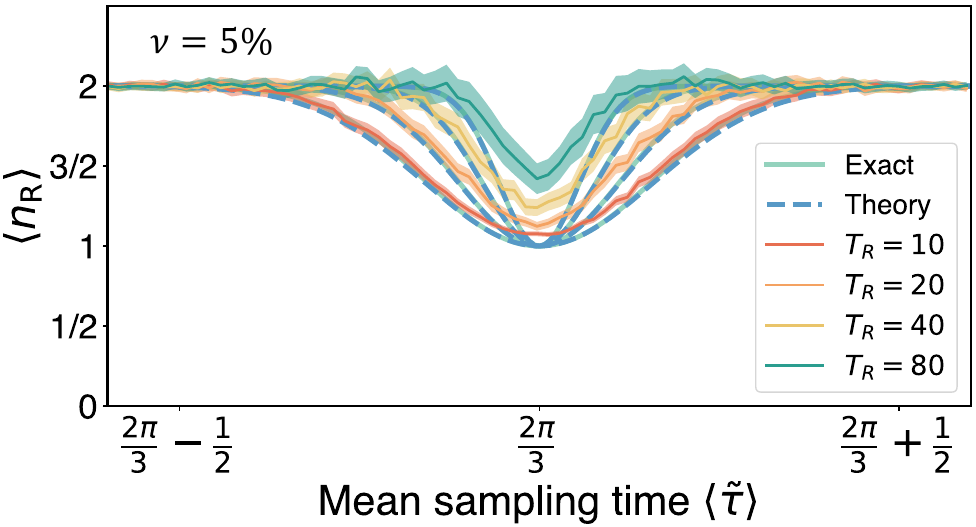}
\end{center}
\caption{Mean hitting time versus the mean sampling time $\langle \tilde{\tau} \rangle $, for the three-site ring model,
with a fixed inaccuracy level of \(5\%\) 
in the evolution time \(\tau\), and varying restart time from \(T_R = 10\) to \(T_R = 80\). 
Using Monte Carlo simulation with \(30,000\) realizations, 
we observed that deviations in the resonances intensify
with increasing restart time \(T_R\). 
However, despite these deviations, 
when $\langle \tilde{\tau} \rangle $ is either small or large, 
namely when $\langle \tilde{\tau} \rangle $ is tuned far from the resonance, 
we see that randomness of $\tau$ is of no consequence, 
and the topological winding number $2$, is robust.
}
\label{fig5}
\end{figure}

\subsection{Randomness in the sampling time $\tau$}
For the randomness in the evolution time $\tau$,
we employed the Monte Carlo method to explore the impact of fluctuations of $\tau$
on the uncertainty relation.
We denote the actual evolution time in experiments by
$\tilde{\tau}$, and it is a uniformly random variable within the range 
$[\tau(1-\nu), \tau(1+\nu)]$.
We will vary the value of $\nu$ from $0.05$ to $0.3$ 
corresponding to \(5\%\) to \(30\%\) inaccuracy levels. 
Here we use the three-site ring model, which was used in our IBM experiment. 
We also study the benzene ring model.
See Figure~\ref{fig3}-\ref{fig5} for numerical results, 

The procedures for the Monte Carlo method,
used to produce Figures \ref{fig3},\ref{fig4},\ref{fig5},
are the following:
%
\begin{enumerate}[(i)]
    \item   {\em Initialization of the quantum walker:} 
            The quantum walker is initially 
            evolved from a predefined state in accordance with 
            the Schrödinger equation 
            This evolution occurs over a time duration, $\tilde{\tau}_1$, 
            which is a uniformly random variable within the range 
            $[\tau(1-\nu), \tau(1+\nu)]$.
    \item   {\em Random coin tossing for detection assessment:} 
            A random variable, referred to as a ``coin'', is generated. 
            This variable is uniformly distributed within the interval $[0, 1]$. 
            The purpose of the coin is to ascertain whether the quantum walker 
            is detected following the initial state's evolution. 
            This determination is made by comparing the coin's value 
            with the detection probability, 
            which is derived from the unitary evolution.
    \item   {\em Non-detection and state modification:} 
            If the coin value falls below the computed detection probability,
            we are done and the hitting time is $1$.
            If the coin value exceeds the computed detection probability, 
            it signifies that the walker remains undetected. 
            In this case, the amplitude at the target site $\ket{0}$ is erased, 
            and the wave vector is renormalized.
            Subsequently, the single-site-erased wave vector undergoes unitary evolution for a duration, $\tilde{\tau}_2$. 
            Notably, $\tilde{\tau}_2$ is an independent and identically distributed 
            (i.i.d.) random value, akin to $\tilde{\tau}_1$. 
            The objective is to compute the probability of detection 
            at the time $t = \tilde{\tau}_1 + \tilde{\tau}_2$.
    \item   {\em Repeated detection attempts: }
            Post the initial non-detection, a second i.i.d. coin is generated 
            and compared with the newly computed detection probability 
            to decide if the walker is detected at this stage, 
            as in the step (iii).
    \item   {\em Criteria for repetition termination under sharp restart:} 
            The process iterates until the coin value is less than 
            the computed probability of detection, 
            marking the end of a repetition cycle. 
            Alternatively, if the process extends up to 
            a preset fixed restart step, $T_R$ 
            (i.e. after a cumulative time of $t=\tilde{\tau}_{1}+\tilde{\tau}_2+\cdots+\tilde{\tau}_{T_R}$),
            and the walker remains undetected, 
            the entire procedure recommences from the initial state, 
            repeating the procedures (i)-(v).
    \item   {\em Conditional/restarted hitting time calculation: }
            Once the system is detected in the target state, for the first time, we are done.
            The number of all preceding unsuccessful attempts, 
            plus $1$, is recorded as the fist-detection time, 
            or the hitting time under restarts, $n_R$.
            For the conditional mean, 
            we need to discard all data where no detection occurs before each restart,
            namely, only the outcome sequences containing ``yes'' are retained
            (as explained in the main text).
    \item   {\em Realizations and expected value determination: }
            The aforementioned procedures, executed for obtaining 
            a single value of the hitting time under $T_R$-step restarts, 
            is called a single realization. 
            To ascertain $\expval{n}_\text{Con}$ or $\expval{n_R}$ as a function of $\tau$,
            large number of realizations are conducted for each value of $\tau$.
\end{enumerate}

Our results indicate that at \(\nu=5\%\), 
the uncertainty relation exhibits minimal change in the mean recurrence time, 
as demonstrated in Figure \ref{fig3}, and only a slight deviation in the mean. 
As the fluctuations of $\tau$ increase, 
these deviations become progressively more pronounced. 
Notably, at an inaccuracy level of \(30\%\), 
the resonances are completely obliterated, 
effectively disrupting the uncertainty relations 
due to the stochastic nature of the evolution time, \(\tau\).
Furthermore, our analysis reveals that such fluctuations of $\tau$ 
does not affect the topological number, 
which is $2$ in this case, underscoring its robustness. 
This resilience may represent a form of topological self-protection.
A similar phenomenon has also been observed in the benzene-type ring model, 
as illustrated in Figure~\ref{fig4}.

We further investigate the scenario with a constant inaccuracy level of \(5\%\) 
and a variable restart time, \(T_R\), 
to examine the influence of increasing \(T_R\) on the system dynamics, 
as illustrated in Figure~\ref{fig5}. 
Our observations reveal that the deviation from the results for ideal cases
(without noise) increases when \(T_R\) grows.
Moreover, at the exact point $\tau=2\pi/3$, 
the mean hitting times depart from $w=1$ of the precise-$\tau$ process. 
Despite these changes, the topological number far from the resonance remains stable, 
underscoring its robustness against variations in the sampling time.

\subsection{Randomness in the restart time $T_R$}
\begin{figure}[t]
\begin{center}
\includegraphics[width=0.45\linewidth]{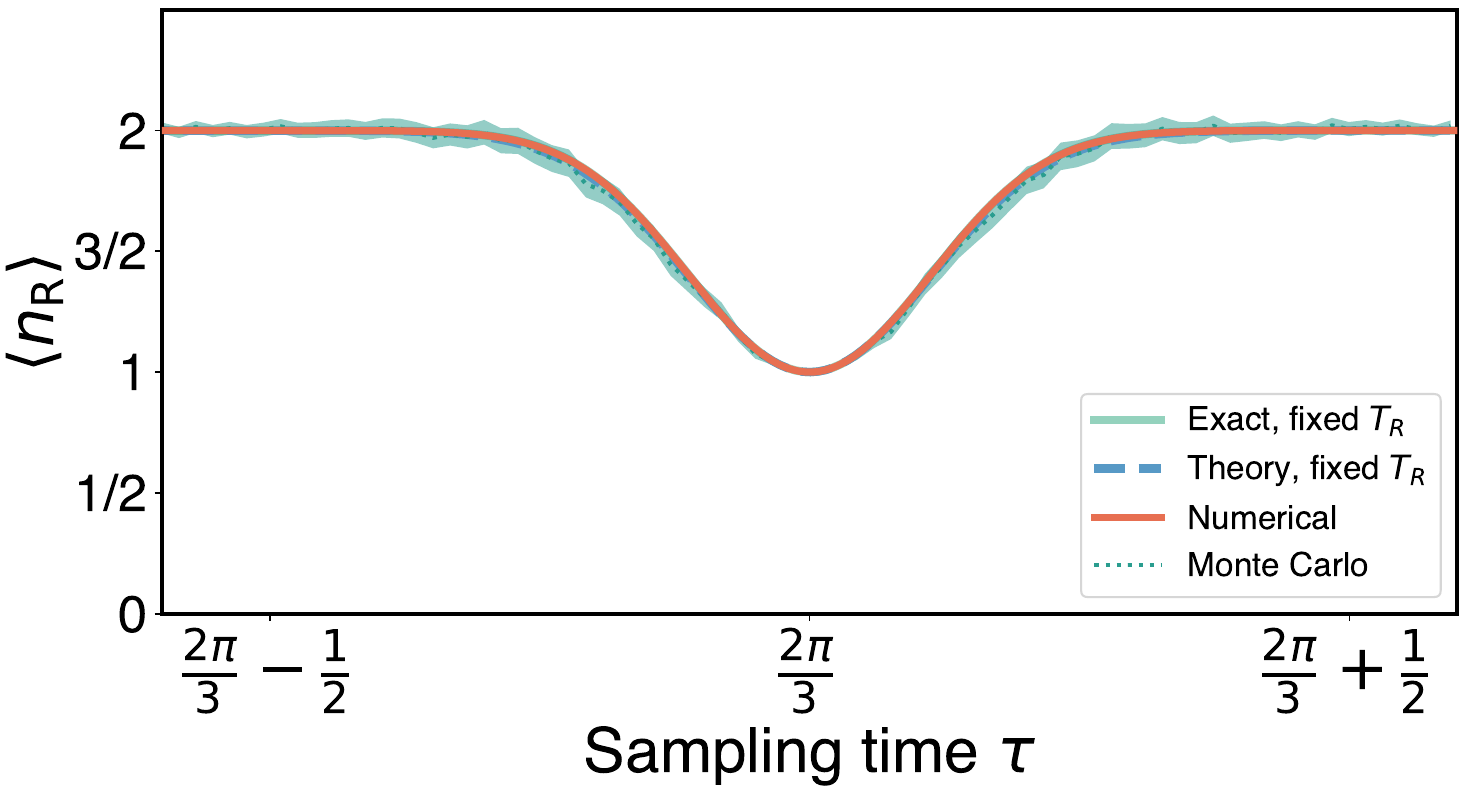}
\end{center}
\caption{ Mean hitting time for the three-site ring model with random \(T_R\), 
where \(T_R\) is drawn from a narrow distribution, 
such that the probabilities of $T_R$ being \(19\), \(20\), and \(21\) 
are \({1}/{4}\), \({1}/{2}\), and \({1}/{4}\), respectively. 
Here we used the mean of $T_R$ equal to $20$, motivated by our quantum computer experiments.
Our findings reveal that this randomness in \(T_R\) 
has a negligible impact on the outcomes. 
Exact results for the restarted mean with fixed $T_R$
are obtained with equations (2-4) in the main text, 
numerical results are calculated with equation (\ref{eqsm15a}), 
and the Monte Carlo simulations are conducted with {\em Python}.
}
\label{fig6}
\end{figure}

Recall that previously, we recorded $T_R$ times, 
which is the duration of the experiment in units of $\tau$. 
Clearly, in common situations with mid-circuit measurements on quantum computers,
this number is fixed 
since experimentalists can typically control and measure the duration of an experiment. 
But in the literature of stochastic restarts the randomness of this variable 
is also considered for classical restart processes.
We will now investigate how the randomness in $T_R$ affects our uncertainty relation.
We assume the restart time $T_R$ assigned to three values,
$19$, $20$ and $21$, with probability $1/4$, $1/2$ and $1/4$, respectively 
(the mean of $T_R$ is still $20$, motivated by our quantum computer experiments).
%
We computed both exact numerical results 
(see the formulas below), 
and simulated results with Monte Carlo methods, as shown in Figure~\ref{fig6}. 
Our analysis reveals that, in each case, 
the randomness in \(T_R\) exerts negligible impact 
on both the uncertainty relations and the stability of the topological number.

Our initial choice of distribution of $T_R$ was rather narrow, 
we therefore also studied the case when $T_R$ is Poisson distributed. 
We have found that also in this case, the effect of randomness in $T_R$ is negligible.
The reason is the following: 
the mean of  $T_R$ was $20$, similar to our IBM experiments. 
In this case, the Poisson distribution is roughly symmetric around its mean, 
similar to a normal distribution.
The important issue is that when $T_R$ is fixed, 
the location of the resonance $\tau$ is independent of $T_R$ and further, 
the width of the resonance is inversely proportional to $T_R$.
Hence, we expect that for a distribution of  $T_R$ symmetric around the mean 
(again, like the normal distribution or a tent distribution), 
the effects of randomness of $T_R$ are negligible. 
For non-symmetric distributions of $T_R$, other effects are expected.

Therefore, to summarize, 
for symmetric distributions of $T_R$, the peak of the distribution is located on the mean,
the time-energy uncertainty relation does not change considerably 
if compared with a theory for which $T_R$ is fixed.
And for fluctuations of the sampling time $\tau$,
our analysis reveals that,
when $T_R$ is not too large,
the time-energy uncertainty relation is not significantly affected.
But the resonance is diminishing when $T_R$ is increased for fixed width of the distribution of $\tau$.
At the same time, the topological number far from resonance is very robust 
to the fluctuations of $\tau$.

\begin{figure}[ht]
\begin{center}
\includegraphics[width=0.45\linewidth]{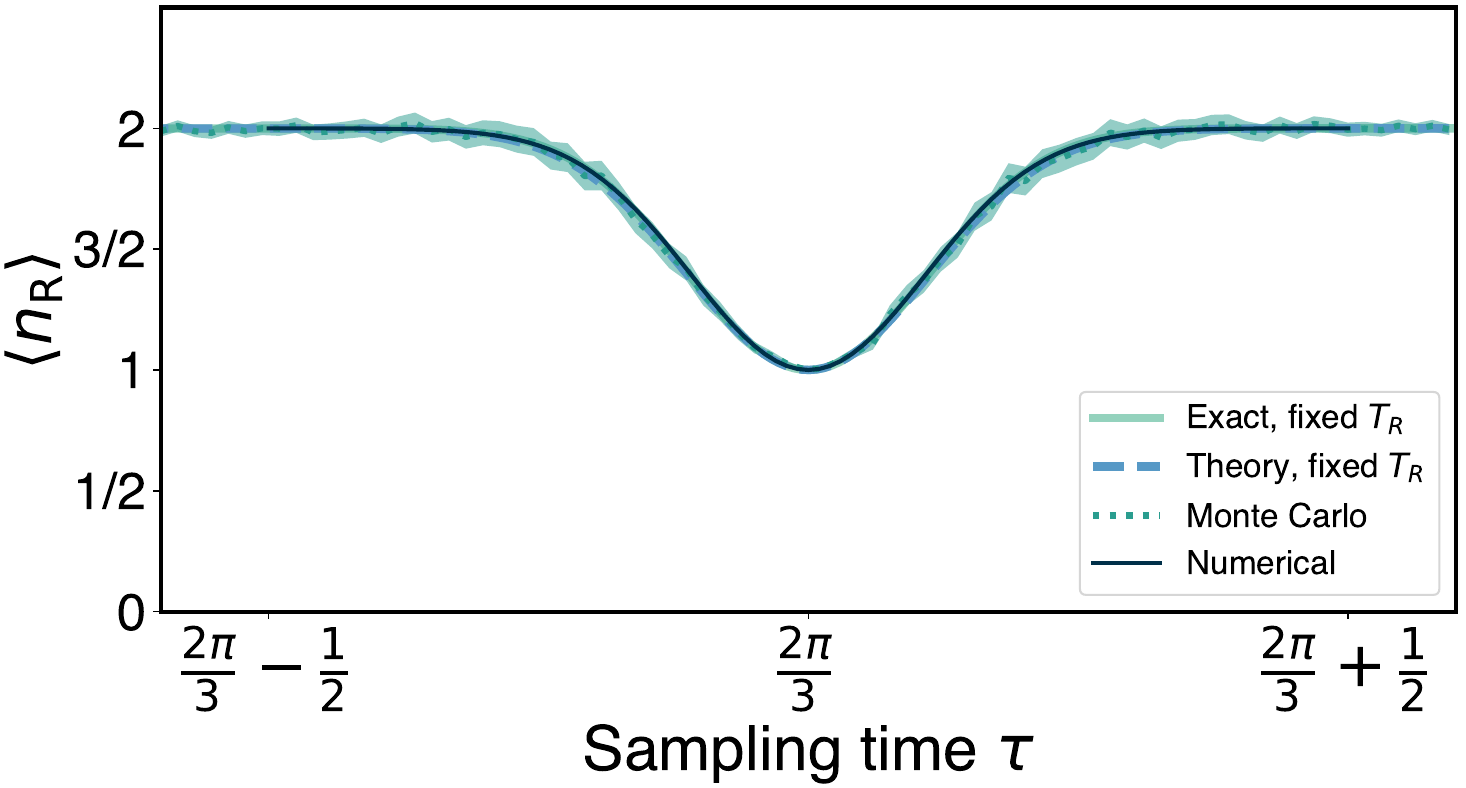}
\end{center}
\caption{Mean hitting time for the three-site ring model 
with Poissonian distributed $T_R$,
and we used $\expval{T_R}=20$.
There appears no deviation from the fixed-$T_R$ case, 
for which the exact results are obtained 
with equations (2-4) in the main text, 
and the theory is computed with equation (7) in the main text.
The numerical results for Poissonian distributed $T_R$ 
are calculated with equation (\ref{eqpoi}) below, 
and the Monte Carlo simulations are conducted with $30,000$ realizations.
}
\label{fig:poi}
\end{figure}
%
%

{\bf Formulas for random restart time.}
The numerical results for random $T_R$ are calculated using the following formula: 
\begin{equation}\label{rtr}
\begin{split}
    \expval{n_R} &=
    {
    \sum_{k=1}^\infty k P(T_R=k) 
    (1-\sum_{n=1}^{k-1} F_n) 
    +
    \sum_{n=1}^\infty nF_n \sum_{k=n+1}^\infty P(T_R=k)
    \over 
    \sum_{n=1}^\infty F_n \sum_{k=n}^\infty P(T_R=k)
    }.
\end{split}
\end{equation}
$P(T_R=k)$ is the probability that restart occurs after $k$ attempts of measurements,
and $F_n$ is the probability of detecting the system 
at the $n$th measurement for the first time, 
in the absence of restarts. 
These basic probabilities are found using equation (2) in the main text.
In equation (\ref{rtr}), 
we employed the general framework proposed by Pal and Reuveni \cite{Shlomi2017,Pal2021},
which states that the mean hitting time with a general distribution of $T_R$ is 
%
\begin{equation}\label{eqsm15}
    \expval{n_R} = {
                    \expval{\text{min} (n, T_R)} 
                    \over
                    P(n\le T_R)
                    }
                    ,
\end{equation}
where $n$ is the first hitting time in the absence of restart, 
and the numerator means the expectation of the minimum 
of $n$ and the random restart time $T_R$.
We note that
\begin{equation}\label{eqsm15a}
\begin{aligned}
    \expval{\text{min} (n, T_R)} 
    =&  \sum_{k=1}^\infty k P(T_R=k) 
        \left( 1-\sum_{n=1}^{k-1} F_n \right) 
        +
        \sum_{n=1}^\infty n F_n \sum_{k=n+1}^\infty P(T_R=k), \\
    P(n\le T_R) =& \sum_{n=1}^\infty F_n \sum_{k=n}^\infty P(T_R=k).
\end{aligned}
\end{equation}
%
Here we used the normalization of $P(T_R=k)$, i.e. $\sum_{k=1}^\infty P(T_R=k) = 1$.
And for the aforementioned distributions of $T_R$ on a finite range, 
the upper limit of the sum associated with $P(T_R=k)$ will be truncated to the largest value of $T_R$.
For the Poisson distribution of $T_R$,
equation (\ref{rtr}) becomes \cite{Pal2021,Ruoyu2023}
\begin{equation}\label{eqpoi}
    \expval{n_R}_{\rm Pois} 
    = 
    { 
    1+\lambda 
    - \sum_{n=1}^{\infty} F_n \sum_{k=n+1}^{\infty}(k-n) 
    \frac{e^{-\lambda} \lambda^{k-1}}{(k-1)!} 
    \over 
    \sum_{n=1}^{\infty} F_n \sum_{k=n}^{\infty} 
    \frac{e^{-\lambda} \lambda^{k-1}}{(k-1)!}
    }, 
\end{equation}
%
%
Here the parameter $\lambda = \expval{T_R}-1$. 
In Figures \ref{fig6},\ref{fig:poi}, 
we utilized equation (\ref{eqpoi}) for the Poisson case, 
and (\ref{rtr}) for the tent-like distribution of $T_R$, 
to generate the ``Numerical'' results.

\section*{5. Implementation on a quantum computer}
The three-site tight-binding Hamiltonian (equation (16) in the main text with $L=3$) is encoded by the qubit Hamiltonian:

\begin{eqnarray*}
    H = -\frac{1}{2} (\sigma_{x, 1}+\sigma_{x, 2}+\sigma_{z, 1} \sigma_{x, 2}+\sigma_{x, 1} \sigma_{z, 2}+\sigma_{x, 1} \sigma_{x, 2}+\sigma_{y, 1}  \sigma_{y, 2} \nonumber)
\end{eqnarray*}
where $\sigma_x$, $\sigma_y$ and $\sigma_z$ are the Pauli matrices. 
The Hamiltonian $H$ defines two disconnected subspaces, 
the first composed of the states $\ket{00},\ket{01},\ket{10}$ 
and the second from $\ket{11}$. 
In our scheme, the state $\ket{11}$ is not used.
Hence, we use the following mapping between the qubit and spatial states representation:
$\ket{01} \rightarrow \ket{0}, \ket{10} \rightarrow \ket{2} \ \mbox{and} \
\ket{00} \to \ket{1}$. 
The unitary evolution operator $U(\tau) = \exp(-i H \tau)$, 
must be constructed on a quantum computer as a product of elementary gate operators.
We explain how to perform the measurements and how to construct an efficient
unitary. 

\begin{figure}[h]
\centering
\includegraphics[width=0.45\textwidth]{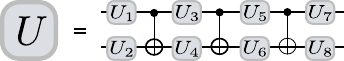}
\caption{Decomposition of the Unitary}
\end{figure}
%

We define the two-qubit unitary transformation
$U(\tau) = \exp( - i H \tau)$ using Cartan's decomposition \cite{Vidal2004},
namely with three CNOT gates and single unitary gates (see sketch). 
%
Importantly, this allows us to vary $\tau$ in
simulations without much computational cost. 
For larger systems, one would have to use other methods to model the unitary, 
namely trotterization technique.

\begin{figure}[h]
\centering
\includegraphics[width=0.45\textwidth]{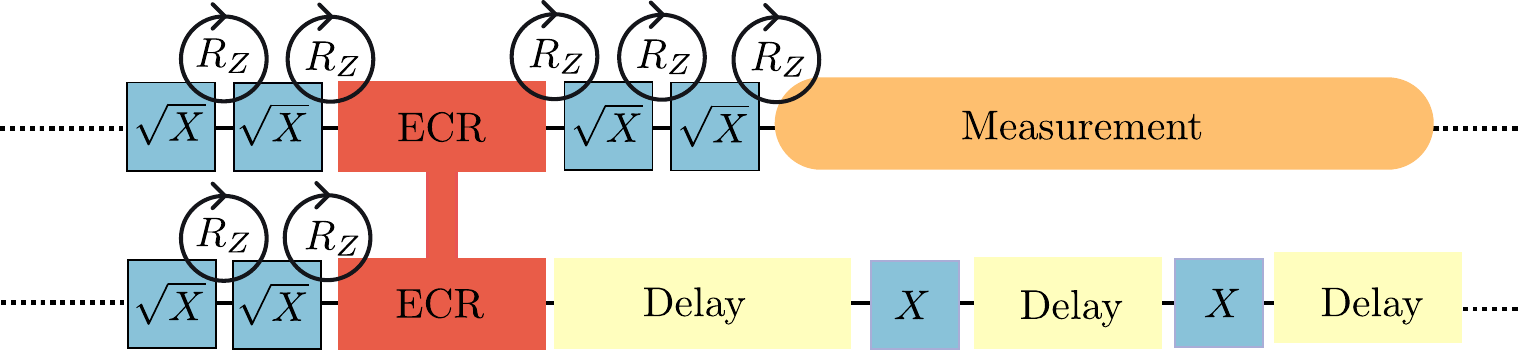}
\caption{{ The schematic timeline for qubit gates in the quantum computer}}
\label{dd}
\end{figure}

In our study, we employ localized single-site measurements, as integrated in the IBM computer toolbox, to detect state $\ket{0}$ without distinguishing states $\ket{1}$ and $\ket{2}$, as mentioned in the main text. 
As an error suppression strategy, we are using dynamical decoupling and inserting two XX-gates on the qubit which is not measured to keep it coherent. 
The schematic timeline is given by Fig.~\ref{dd}.

%


\end{document}